\documentclass[twocolumn,pra]{revtex4}
\usepackage{psfrag}
\usepackage{amsmath}
\usepackage{amssymb}
\usepackage{epsfig}
\usepackage{longtable}

\usepackage[colorlinks=true,linkcolor=blue,citecolor=blue]{hyperref}

\newcommand{\bra}[1]{\ensuremath{\langle#1|}}
\newcommand{\ket}[1]{\ensuremath{|#1\rangle}}
\newcommand{\braket}[2]{ \ensuremath{ \langle #1|#2\rangle}}
\newcommand{\comm}[2]{\left [ #1, #2 \right]}
\newcommand{\be}{\begin{equation}}
\newcommand{\ee}{\end{equation}}
\newcommand{\avg}[1]{\ensuremath{\langle #1 \rangle}}

\newcommand{\im}{\text{i}}

\newcommand{\ie}{{\it i.e.}}
\newcommand{\cf}{{\it cf.}}

\newcommand{\etal}{{\it et al.}}

\newcommand{\eqcite}[1]{Eq.~\eqref{#1}}

\newcommand{\xx}{\bold{x}}

\newcommand{\rr}{\bold{r}}

\newcommand{\nn}{\bold{n}}

\newcommand{\xop}{\hat x}
\newcommand{\xtild}{\tilde x}
\newcommand{\pop}{\hat p}
\newcommand{\aop}{\hat a}
\newcommand{\adop}{\hat a^{\dagger}}
\newcommand{\ain}{\hat a_{\text{in}}}
\newcommand{\aout}{\hat a_{\text{out}}}

\newcommand{\adout}{\hat a^{\dagger}_{\text{out}}}
\newcommand{\Pout}{\hat P^{\text{L}}_{\text{out}}}
\newcommand{\Pin}{\hat P^{\text{L}}_{\text{in}}}

\newcommand{\Xin}{\hat X^{\text{L}}_{\text{in}}}

\newcommand{\acron}{MERID}

\begin{document}

\title{Quantum Superposition of Massive Objects and Collapse Models}

\author{Oriol Romero-Isart}
\affiliation{Max-Planck-Institut f\"ur Quantenoptik,
Hans-Kopfermann-Str. 1,
D-85748, Garching, Germany}

\begin{abstract}
We analyze the requirements to test some of the most paradigmatic collapse models with a protocol that prepares quantum superpositions of massive objects. This consists of coherently expanding the wave function of a ground-state-cooled mechanical resonator, performing a squared position measurement that acts as a double slit, and observing interference after further evolution. The analysis is performed in a general framework and takes into account only unavoidable sources of decoherence: blackbody radiation and scattering of environmental particles. We also discuss the limitations imposed by the experimental implementation of this protocol using cavity quantum optomechanics with levitating dielectric nanospheres. 
\end{abstract}

\maketitle

\section{Introduction}

In the last decades, seminal experiments have demonstrated that massive objects can be prepared in spatial superpositions of the order of its size. This has been realized with electrons~\cite{Davisson1927}, neutrons~\cite{Halban1936}, atoms and dimers~\cite{Estermann1930}, small van der Waals clusters~\cite{Schollkopf1994}, fullerenes~\cite{Arndt1999}, and even with organic molecules containing up to 400 atoms~\cite{Gerlich2011}. These experiments are designed to observe the interference of matter-waves after passing, in essence, through a Young's double slit. The possibility of observing these quantum pheonomena with yet larger objects is extremely challenging. This is due to the great quantum control and isolation from the environment that these experiments require. 

More recently, the field of cavity quantum electro- and optomechanics~\cite{Kippenberg2008,Aspelmeyer2008,Marquardt2009,Favero2009,Aspelmeyer2010} has opened the pathway to bring much more massive objects to the quantum regime, namely, objects containing \emph{billions} of atoms, thereby improving the previous benchmark by many orders of magnitude. This allows to explore the physics of a completely new parameter regime. 
A first step towards this direction has been realized in Refs.~\cite{O'Connell2010,Teufel2011,Chan2011}, where ground state cooling of mechanical resonators at the nano- and microscale has been achieved. Additionally, various researchers have proposed to exploit the coherent coupling of the mechanical resonator with single photons or qubits to create quantum superpositions, see for instance~\cite{Marshall2003,Romero-Isart2010b}. In these proposals, the superposition of the mechanical motion state is, typically, of the form $\ket{0} + \ket{1}$, where $\ket{0}$ and $\ket{1}$ are respectively the ground  state and the first excited state of the harmonic potential. In these states, the position is delocalized over distances of the order of the zero point motion, \ie~$x_{0}=\sqrt{\hbar/(2 m \omega)}$, where $m$ is the mass of the object and $\omega$ the frequency of the harmonic potential. Within the megahertz regime, objects containing $n_{\text{at}}$ atoms are delocalized over distances of the order of $10^{-7} n_{\text{at}}^{-1/2}$ m, which is subatomic for objects containing billions of atoms. This is  in contrast with matter-wave experiments, where, despite the fact that objects have ``only'' hundreds of atoms, they can be delocalized over distances larger than their size.

Remarkably, these experiments might be applied to the service of a very fundamental goal, namely, the exploration of the limits of quantum mechanics predicted by several collapse models~\cite{Diosi1984,Ghirardi1986,Frenkel1990,Ghirardi1990a,Ellis1992,Penrose1996,A.J.Leggett2002,Bassi2003,Diosi2007, Adler2009}. The common idea of these models is the conjecture that the Schr\"odinger equation is an approximation of a more fundamental equation, which breaks down when objects above a critical mass are delocalized over a critical distance. This prediction is very difficult to confront because of the following argument: standard decoherence~\cite{Joos2003,Schlosshauer2007}, described within quantum mechanics, also predicts the impossibility to delocalize large objects due to the interaction with the environment; thus, this masks the effects of collapse models. This poses a major challenge to corroborate collapse models, as the effects predicted by these must stand alone from decoherence processes and be exposed to potential falsification.
This leads to the central questions of this work: How challenging is it to test collapse models while also taking into account unavoidable sources of decoherence? Is it preferable to have small objects delocalized over large distances, as in matter-wave experiments, or rather large objects delocalized over small distances, as in experiments with mechanical resonators?

The aim of this paper is to address the latter questions by analyzing a prototypical experiment that bridges approaches from quantum-mechanical-resonators and matter-wave interferometry. This experiment relies, on the one hand, on techniques of cavity electro-optomechanics to prepare a mechanical resonator in the ground state of its harmonic potential. On the other hand, the experiment mimics matter-wave interferometry,  as the ground-state-cooled mechanical resonator is released from the harmonic trap in such a way that it coherently delocalizes over distances much larger than its zero point motion $x_{0}$. 
A subsequent measurement of the squared position, which is to be realized using techniques of quantum-mechanical resonators, collapses the state into a superposition of different spatial locations, thereby acting as a Young's double slit. Finally, the subsequent free evolution generates an interference pattern. We remark that the implementation of this experiment using cavity quantum optomechanics with optically levitating dielectric nanospheres~\cite{Romero-Isart2010b, Chang2010, Barker2010a, Romero-Isart2011} has been recently proposed in~\cite{Romero-Isart2011c}. 
The present article analyses this proposal with a broader scope, namely, it studies the effect of some of the most paradigmatic collapse models together with unavoidable sources of decoherence. This allows us to obtain the environmental conditions, masses of the objects, and delocalization distances where collapse models can be falsified. These conditions are general and will be common to any physical implementation of the protocol.

This article is organized as follows: in Sec.~\ref{sec:doubleslit} we introduce and analyze the quantum-mechanical resonator double slit experiment in a general fashion, neglecting decoherence and without specifying the experimental implementation. The effect of unavoidable sources of decoherence, such as blackbody radiation and scattering of environmental particles, will be the subject of Sec.~\ref{Sec:StandardDecoherence}. The effects of several collapse models  in this experiment are discussed in Sec.~\ref{Sec:CollapseModels}, where we also obtain the parameter regime needed to confront them. Finally, in Sec.~\ref{sec:OMdoubleslit}, we study the restrictions imposed by an implementation of this experiment using cavity optomechanics with optically levitating dielectric nanospheres. We draw our conclusions and provide further directions in Sec.~\ref{sec:conclusions}.

\section{Mechanical Resonator Interference in a Double Slit} \label{sec:doubleslit}

\begin{figure}
\begin{center}
\includegraphics[width=\linewidth]{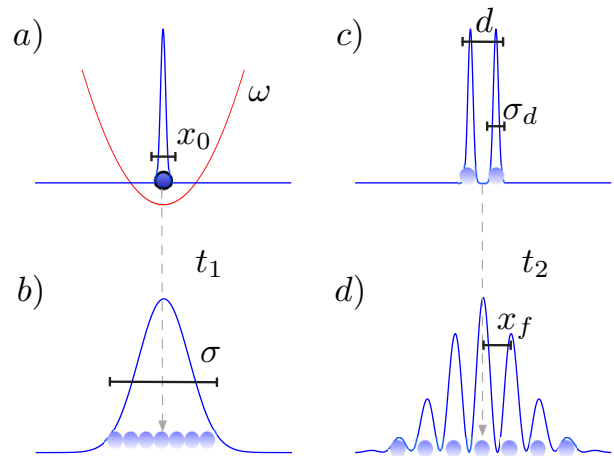}
\caption{(color online) Schematic illustration of \acron. a) A sphere of mass $m$ is harmonically trapped, with frequency $\omega$, and cooled into the ground state. The zero point motion is given by $x_{0}$. b) The trap is switched off and the wave function expands freely during some time of flight $t_{1}$. At this time, the width of the wave function is given by $\sigma$. c) A squared position measurement is performed such that the wave function collapses into a superposition of two wave packets of size $\sigma_{d}$ separated by a distance $d$. Both $\sigma_{d}$ and $d$ depend on the measurement outcome. d) The superposition state evolves freely during a second time of flight $t_{2}$. An interference pattern is formed with peaks separated by $x_{f}$. }
\label{Fig:Scheme}
\end{center}
\end{figure}

In this section we analyze a protocol that merges techniques and insights from quantum-mechanical resonators and matter-wave interferometry; we call it MEchanical Resonator Interference in a Double slit (\acron). We analyze it without taking into account standard decoherence (\cf~Sec.~\ref{Sec:StandardDecoherence}) and without specifying its experimental implementation (\cf~Sec.~\ref{sec:OMdoubleslit}). The \acron~is realized by applying the following steps, see Fig.~\ref{Fig:Scheme}: 
\begin{enumerate}
\item\label{step1} Prepare a mechanical resonator. For instance, trap a massive object of mass $m$, which is typically a sphere, into an harmonic potential with trap frequency $\omega$. 
\item\label{step2}  Cool the center-of-mass along one direction, say $\xop$, to, ideally, the ground state of the harmonic potential.
\item\label{step3}  Switch off the harmonic trap and let the wave function expand freely during some ``time of flight'' $t_{1}$.
\item\label{step4}  Perform a measurement of $\xop^{2}$, that is, of the squared position of the cooled degree of freedom. This measurement acts as a Young's double slit since, given the outcome $x^{2}$, the state collapses into a superposition of being at $+x$ and at $-x$. The mechanical resonator is thus prepared in a spatial superposition separated by a distance $d=2 |x|$.
\item\label{step5}  Let the state evolve freely during a second time of flight $t_{2}$.
\item\label{step6}  Perform a measurement of the center-of-mass position $\xop$.
\item\label{step7}   Repeat the experiment and collect the data for each double slit distance $d$, corresponding to the result of the squared position measurement. An interference pattern in the final position measurement is unveiled for each $d$.    
\end{enumerate}

\subsection{Steps \ref{step1} and \ref{step2}: Cooled initial state}

These steps consist in preparing the object's the center-of-mass motion along $x$ in the ground state of an harmonic potential with a trapping frequency $\omega$, see Fig.~\ref{Fig:Scheme}a. In the ideal case, the wave function is given by
\be \label{eq:GS}
\braket{x}{0}= \frac{1}{[2 \pi x^{2}_{0}]^{1/4}}\exp \left[- \frac{x^{2}}{4 x^{2}_{0}} \right],
\ee
where $x_{0}=\sqrt{\hbar/(2 m \omega)}$ is the ground state size and $m$ is the mass of the object. In realistic situations, the initial state is given by a thermal state with mean occupation number $\bar n = (\exp[ \beta \hbar \omega]-1)^{-1}$ (where $\beta^{-1}=k_{b}T$,  $k_{b}$ is the Boltzmann constant, and $T$ the effective one dimensional center-of-mass temperature), which can be written in the Fock basis as
\be \label{eq:thermal}
\hat \rho(0)= \sum_{n=0}^{\infty} \frac{\bar n^{n}}{(1+\bar n)^{n+1}} \ket{n} \bra{n}.
\ee
This state has the following moments
$\avg{\xop^{2}(0)} = (2 \bar n +1) x^{2}_{0}$,  
$\avg{\pop^{2}(0)} = (2 \bar n +1) \hbar^{2}/(4 x^{2}_{0})$, 
and $\avg{\comm{\xop(0)}{ \pop(0)}_{+}}=0$.
We do not discuss here how cooling is performed experimentally, see however~\cite{Romero-Isart2010b, Chang2010, Barker2010a,Romero-Isart2011,Barker2010} for optomechanical cooling techniques~\cite{Marquardt2007,Wilson-Rae2007,Genes2008} applied to optically levitating nanospheres.

\subsection{Step \ref{step3}: Expansion}
This step (see Fig.~\ref{Fig:Scheme}b) consists in switching off the trap and letting the wave function evolve freely, that is, it evolves with the unitary time evolution $\hat U_{0}(t)=\exp[- \im \hat H_{0} t /\hbar]$, where $\hat H_{0}= \pop^{2}/(2m)$. Considering the initial state to be the pure ground state, the state after some time $t_{1}$ is given by
\be \label{eq:psit1}
\bra{x} \hat U_{0}(t_{1}) \ket{0} =  \frac{1}{[2 \pi \sigma^{2}]^{1/4}} \exp \left[- \frac{ x^{2}}{ 4 \sigma^{2} } +\im   \phi_{\text{tof}}   \frac{x^{2}}{\sigma^{2}}  \right],
\ee
where $\sigma^{2} =x^{2}_{0} (1 + t_{1}^{2} \omega^{2})$
is the size of the expanded wavefunction, and
$\phi_{\text{tof}}=\omega t_{1}/4$
is the global phase accumulated during the free evolution.

\subsection{Step \ref{step4}: Double slit} \label{sec:DS}
\begin{figure}
\begin{center}
\includegraphics[width=\linewidth]{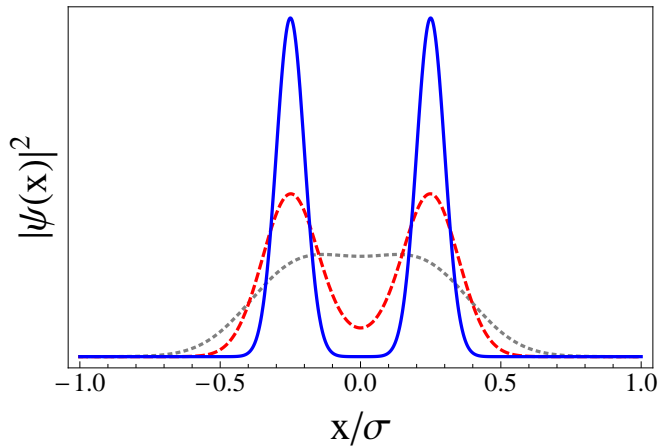}
\caption{(color online) $|\psi(x)|^{2}=|\braket{x}{\psi}|^{2}$, see \eqcite{eq:DSfunction}, is plotted for $d=\sigma/2$ and measurement strength $\chi=6$ (dotted gray), $\chi=10$ (dashed red), and $\chi=20$ (solid blue). }
\label{Fig:DoubleSlit}
\end{center}
\end{figure}

In this step a squared position measurement of the state at time $t=t_{1}$ is performed, see \eqcite{eq:psit1}, such that the state collapses into
\be \label{eq:DSfunction}
\ket{\psi}\equiv \frac{\hat{\mathcal{M}}_{d} \hat U_{0}(t_{1}) \ket{0}}{ || \hat{\mathcal{M}}_{d} \hat U_{0}(t_{1}) \ket{0} ||}.
\ee
The measurement operator, $\mathcal{M}_{d}$, is assumed to have the following form
\be \label{eq:mesop}
\begin{split}
\hat{\mathcal{M}}_{d} =& \exp\left[\im \phi_{\text{ds}} (\xop/\sigma)^{2} \right] \times \\
& \times \left\{ \exp \left[ - \frac{\left(\xop - \frac{d}{2}\right)^{2}}{4 \sigma_{d}^{2}}  \right] + \exp \left[ - \frac{\left(\xop + \frac{d}{2}\right)^{2}}{4 \sigma_{d}^{2}} \right] \right\}.
\end{split}
\ee
This measurement has the potential to prepare a quantum superposition of Gaussian wavefunctions of width $\sigma_{d}$ separated by a distance $d$, with an added global phase that we discuss below. The state $\ket{\psi}$ presents a well-resolved spatial superposition provided that $ d > 2\sigma_{d}$. Also, one requires  $\sqrt{8} \sigma > d$ in order to have a non-negligible probability to obtain the result $d$, that is, in order to ensure  that $|\bra{d/2} \hat U_{0}(t_{1}) \ket{d/2} |^{2}/|\bra{0} \hat U_{0}(t_{1}) \ket{0} |^{2}>\exp[-1]$. We have summarized in Table~\ref{t:summary} all the conditions required to successfully realize \acron~that will be obtained throughout the article. These two obtained here are included as conditions (i) and (ii) by using the definition given below in \eqcite{eq:chi}. Motivated by the optomechanical implementation of this measurement (see Sec.~\ref{sec:OMdoubleslit} where we derive \eqcite{eq:mesop}), let us define a dimensionless parameter independent of the measurement result,  which characterizes the strength  of the measurement, and relates $d$ with $\sigma_{d}$ as
\be \label{eq:chi}
\chi \equiv \frac{\sigma^{2}}{2 \sigma_d d}.
\ee
For a given outcome $d$, the larger the value of $\chi$, the more resolved the superposition. Figure~\ref{Fig:DoubleSlit} shows the position probability distribution of the state of \eqcite{eq:DSfunction} with $d=\sigma/2$ for different measurement strengths $\chi$.

\begin{figure}
\begin{center}
\includegraphics[width=\linewidth]{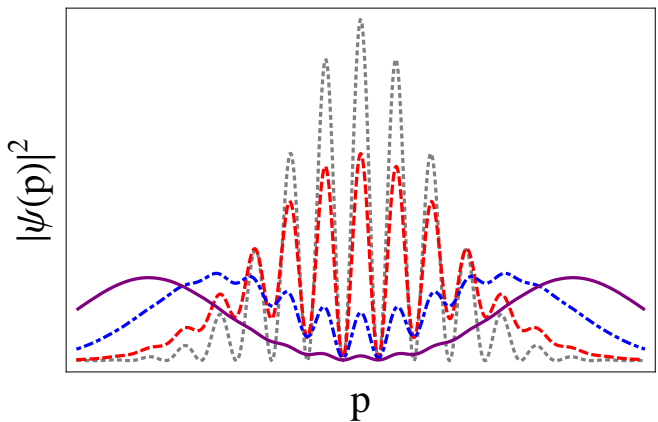}
\caption{(color online) $|\psi(p)|^{2}=|\braket{p}{\psi}|^{2}$, see \eqcite{eq:DSfunction}, is plotted  in arbitrary units for $d=\sigma/2$, measurement strength $\chi=50$, and $\alpha \equiv |\phi_{\text{ds}} + \phi_{\text{tof}}|$ equal to $\alpha=0$ (dotted gray), $\alpha=50$ (dashed red),  $\alpha=100$ (dot-dashed blue), and $\alpha=150$ (solid purple). }
\label{Fig:Phases}
\end{center}
\end{figure}

Finally, note that a global phase $\phi_\text{ds}$ is added during the measurement. This phase, as well as the one accumulated during the time of flight, $\phi_{\text{tof}}$ in \eqcite{eq:psit1}, plays an important role. The condition $|\phi_\text{ds} + \phi_{\text{tof}}| d^{2}/(4 \sigma^{2}) \ll 1$ needs to be fulfilled in order to build the interference of the two wave packets centered at $x=d/2$ and $x=-d/2$. This can be shown by analyzing $\braket{p}{\psi}$, that is \eqcite{eq:DSfunction} in momentum space, for different global phases, see Fig.~\ref{Fig:Phases}. Recall that within free evolution, the probability momentum distribution of a wave function at $t_{1}$, has the same form as the position probability distribution at much latter times since $\xop(t_{1}+t_{2}) \approx \pop(t_{1}) t_{2}/m$. More intuitively, the global phase adds some momentum to the wave packets. Depending on the sign of this phase, the two wave packets either move apart or towards each other. If the momentum given is too large, for the former case they will separate with a velocity faster than their expansion rate, and thus, they will never overlap. For the latter case, they overlap only during the ``collision'', however, at this time, the wave packets have nearly not expanded and the fringes in the interference pattern cannot be resolved, see the discussion below. 

\subsection{Steps \ref{step5} and \ref{step6}: Interference}

These steps consist in measuring the position distribution of the state obtained after letting the system evolve freely during a second time of flight $t_{2}$; this reads
\be \label{eq:finalWF}
\ket{\psi_{f}}\equiv \frac{\hat U_{0}(t_{2})\hat{\mathcal{M}}_{d} \hat U_{0}(t_{1}) \ket{0}}{ || \hat U_{0}(t_{2}) \hat{\mathcal{M}}_{d} \hat U_{0}(t_{1}) \ket{0} ||}.
\ee
The state $\ket{\psi_{f}}$ presents interference peaks separated by a distance $x_{f}=2 \pi \hbar t_{2}/(md)$ as long as $|\phi_\text{ds} + \phi_{\text{tof}}| d^{2}/(4 \sigma^{2}) \ll 1$. The peaks are clearly visible when the two wave packets overlap, that is, when $d= t_{2}\hbar/(2 \sigma_{d} m)$. Using \eqcite{eq:chi} and $\sigma^{2} \approx x^{2}_{0} t^{2}_{1} \omega^{2}$ (valid at $t_{1} \omega \gg 1$), one obtains an upper bound for $t_{1}$ given by
$t_{1} \lesssim \sqrt{2 t_{2} \chi /\omega}$; this corresponds to condition (iii) in Table~\ref{t:summary}. Another condition is given by the requirement to resolve the interference fringes. Assuming a position resolution of $\delta x$, one requires $x_{f} > \delta x$, which provides an upper bound for the slit distance $d$ given by $d < 2 \pi \hbar t_{2}/(m \delta x)$; this sets condition (iv) in Table~\ref{t:summary}.  The \acron~is finished in step \ref{step7} where the protocol is repeated to obtain a different interference pattern for each double slit length $d$.

\begin{table*}
\begin{center}
\begin{ruledtabular}
    \begin{tabular}{  l | l | l | l  |  }
     & $t_{1}$ & $d$ & $t_{2}$      \\ \hline \hline
i) Highly probably outcome     & - & $ < \sqrt{8}\sigma $ & -       \\ \hline
ii) Superposition peaks resolved     & - & $ > \sigma /\sqrt{\chi} $& -        \\ \hline
iii) Wave packets overlap    &  $ \lesssim \sqrt{2 t_{2} \chi /\omega}$ & - & -     \\ \hline
iv) Fringes can be resolved     & - & $< 2 \pi \hbar t_{2}/(m \delta x)$ & -       \\ \hline
v) Decoh.   expansion  if $d \ll 2a$ & Optimal: $ t_{\text{max}}$& $ < \xi (t_{1}) \le \xi_{\text{max}}$ & -        \\ \hline
vi) Decoh.   expansion   if $d \gg 2a$ & $\ll 1/\gamma $ & $< \xi_{s}(t_{1})$& -  \\ \hline   
vii) Decoh.  interference  if $d \ll 2a$ & - & $ < \sqrt{3/(\Lambda t_{2})}$ & -        \\ \hline
viii) Decoh.  interference   if $d \gg 2a$ & - & - & $ \ll 1/\gamma$  \\ \hline   

ix) Optomechanical implementation     & $\ll  \min\{\sqrt{\kappa/g_{0}},4 g_{0}/\Gamma^{0}_{\text{sc}} \}/\omega$ &$< \xi_{s}(t_{1})$&-       
    \end{tabular}
\end{ruledtabular}
\caption{\label{t:summary}. Summary of the restrictions on the expansion time $t_{1}$, the superposition size $d$, and the evolution time forming the interference pattern $t_{2}$ of the \acron~experiment. Conditions (i-iv) are discussed in Sec.~\ref{sec:doubleslit} and depend on the measurement strength. Conditions (v-viii) depend on the position-localization decoherence and are derived in Sec.~\ref{Sec:StandardDecoherence}. Finally, condition (ix) is given by the optomechanical implementation of \acron, which is the subject of Sec.~\ref{sec:OMdoubleslit}. Recall the definitions of $ t_{\text{max}}$ in \eqcite{eq:tmax}, of $\xi_{\text{max}}$ in \eqcite{eq:ximax}, and of $\xi_{s}(t_{1})$ in \eqcite{eq:xis}.}
\end{center}
\vspace{-0.6cm}
\end{table*}

\section{Decoherence} \label{Sec:StandardDecoherence}

In the previous Section we obtained conditions (i-iv) in Table~\ref{t:summary} for a successful realization of  \acron. Note that, since $t_{1}$ and $t_{2}$ are unbounded, conditions (i-iv) do in principle allow for the preparation of arbitrarily large superpositions. This is the stage when one has to take into account the effect of decoherence, which is the subject of this Section.  We will concentrate on unavoidable sources of decoherence, that is, on the decoherence caused by the interaction with  environmental massive particles (\cf~Sec.~\ref{sec:gas}), and the effects of blackbody radiation (\cf~Sec.~\ref{sec:bb}). We start in Sec.~\ref{sec:posloc} by analyzing a general form of decoherence called position localization. We derive the limitations that this imposes on the expansion time $t_{1}$ (and therefore the superposition size $d$), as well as to the visibility of the interference pattern. This form of decoherence includes the  standard sources of decoherence  mentioned above as well as the effect of collapse models, which we discuss in Sec.~\ref{Sec:CollapseModels}.  

\subsection{Position-localization decoherence} \label{sec:posloc}

\subsubsection{Master equation}

The main feature of a position-localization decoherence is the exponential decay of position correlations, \ie~$\bra{x}\hat \rho (t) \ket{x'} \propto e^{- \Gamma t} \bra{x}\hat \rho (0) \ket{x'} $, where $\Gamma$ usually depends on $|x-x'|$. This form is common both to the decoherence caused by interaction with the environment~\cite{Joos2003,Schlosshauer2007}  and to the exotic one caused by collapse models~\cite{Diosi1984,Ghirardi1986,Frenkel1990,Ghirardi1990a,Ellis1992,Penrose1996,A.J.Leggett2002,Bassi2003,Diosi2007, Adler2009}.  The qualitative behavior of this source of decoherence is very well described by the following master equation given in the position basis
\be \label{eq:localizationgeneral}
\bra{x}\dot{ {\rho}}(t) \ket{x'} = \frac{\im}{\hbar} \bra{x} [\hat \rho,\hat H] \ket{x'}-\Gamma(x-x') \bra{x} \hat \rho(t) \ket{x'},
\ee
\begin{figure}
\begin{center}
\includegraphics[width=\linewidth]{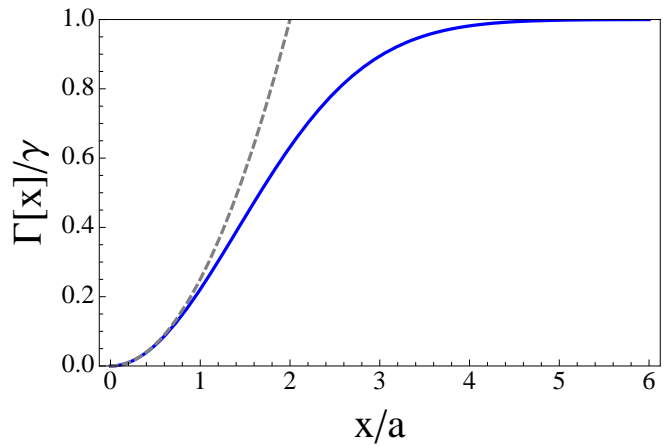}
\caption{(color online) The correlation function $\Gamma(x)$, see \eqcite{eq:Gamma}, is plotted (solid blue line). The short-limit approximation $\Gamma(x)=\Lambda x^{2}$ is also plotted for comparison (dashed gray line).}
\label{Fig:CorrelationFunction}
\end{center}
\end{figure}
where we assume one-dimension for simplicity. The decoherence rate function is defined by
\be \label{eq:Gamma}
\Gamma(x)= \gamma \left(1- \exp \left[ - \frac{x^{2}}{4 a^{2}}  \right] \right).
\ee
This function depends on two parameters: the localization strength $\gamma>0$, which has dimensions of frequency, and the localization distance $a>0$, which has dimensions of length. The value of these parameters depends on the particular source of decoherence, such as gas scattering, black-body radiation, or the one given by collapse models. This simple master equation captures the important feature of position localization decoherence, namely, the saturation behavior
\be
\Gamma(x) \approx \left\{ \begin{array}{ll}
                 \Lambda x^2  
, & x \ll 2a ,\\
              \gamma  
, & x \gg 2 a, \end{array} \right.
\ee
where we have defined the localization parameter $\Lambda \equiv \gamma/(4a^2)$. That is, in the short-distance limit, $|x-x'| \ll 2a$, the position correlations decay  (ignoring the coherent evolution given by the Hamiltonian) as $\bra{x}\hat \rho (t) \ket{x'} \propto e^{- \Lambda |x-x'|^{2} t} \bra{x}\hat \rho (0) \ket{x'} $, such that the decoherence rate depends quadratically on $|x-x'|$. In this limit, \eqcite{eq:localizationgeneral} reads
\be \label{eq:localization} 
\dot \rho(t) = \frac{\im}{\hbar} \comm{\hat \rho(t)}{ \hat H} - \Lambda \comm{\xop}{\comm{\xop}{\hat \rho(t)}}.
\ee
The decoherence rate saturates in the long-distance limit $|x-x'| \gg 2a$. In this regime, the rate is independent of $|x-x'|$ and the position correlations decay as $\bra{x}\hat \rho (t) \ket{x'} \propto e^{- \gamma t} \bra{x}\hat \rho (0) \ket{x'} $. For instance, in Sec.~\ref{sec:gas}, we will see that this is the limit where the wavelength of the particles impinging the object is smaller than the separation of a superposition state, such that a single scattering event resolves the position of the object and provides which-path information. 

In the rest of the article, our strategy will be to approximate each source of position localization decoherence by the simple master equation of the form \eqref{eq:Gamma}, obtaining the localization strength $\gamma$ and distance $a$. Let us therefore analyze the restrictions that the master equation \eqref{eq:Gamma} impose on \acron.

\subsubsection{Solution of the Master equation}

In \acron, different steps assume the free evolution given by the Hamiltonian $\hat H_{0}=\pop^{2}/(2m)$. Hence, we need to obtain the solution of \eqcite{eq:localizationgeneral} for the free dynamics case. This is given by~\cite{Ghirardi1986}
%
\be \label{eq:solution}
\begin{split}
\bra{x} \hat \rho(t) \ket{x'} = &\int_{-\infty}^{\infty}    \frac{dp dy}{2 \pi \hbar} e^{- \im p y /\hbar}   \\ 
& \times \mathcal{F}(p,x-x',t)  \bra{x+y} \hat \rho_{s}(t) \ket{x'+y},
\end{split}
\ee
%
where $ \rho_{s}(t)$ denotes the evolution of the density matrix with the Schr\"odinger equation only, that is, when $\Gamma(x)=0$. The function
\be
{\mathcal{F}}(p,x,t)= e^{-\gamma t} \exp \left[ \gamma \int_{0}^{t} d \tau e^{-  [(x-p \tau /m)/(2 a)]^{2}}  \right]
\ee
takes into account the effects of decoherence. We will use this solution in the following to obtain  different restrictions on \acron.

\subsubsection{Free evolution: coherence length}

We have seen that the step \ref{step4} of \acron~implements a double slit. It is crucial to ensure that the squared position measurement prepares a quantum superposition instead of a statistical mixture. For this to happen, the coherence length of the state before the measurement has to be larger than the slit separation $d$. The coherence length is obtained by analyzing the decay of the position correlation function $\mathcal{C}(x,t) \equiv \bra{x/2} \hat \rho (t) \ket{-x/2} $ as a function of the distance. 

Let us begin by using the solution \eqcite{eq:solution} to obtain the time evolution of the mean values and moments of $\xop$ and $\pop$. It is straightforward to observe that the mean values are not perturbed by the position localization decoherence, that is, $\avg{\xop(t)}=\avg{\xop(t)}_{s}= \avg{\xop(0)}+t \avg{\pop(0)}/m $, $\avg{\pop(t)}=\avg{\pop(t)}_{s}=\avg{\pop(0)}$. However, decoherence does modify the time evolution of the second order moments
\be \label{eq:moments}
\begin{split}
&\avg{\xop^2(t)} = \avg{\xop^2(t)}_{s}+ \frac{2 \Lambda \hbar^{2} }{3 m^2} t^{3}, \\
&\avg{\pop^2(t)} = \avg{\pop^2(t)}_{s}+2 \Lambda \hbar^2 t, \\
&\avg{\comm{\xop(t)}{ \pop(t)}_{+}} = \avg{\comm{\xop(t)}{ \pop(t)}_{+}}_{s}+\frac{2 \Lambda \hbar^{2} t^{2}}{m}. \\
\end{split}
\ee
Here, $\avg{\xop^2(t)}_{s}= \avg{\xop^2(0)}+ \avg{\pop^2(0)}t^{2}/(2m)$, $\avg{\pop^2(t)}_{s}=\avg{\pop^2(0)}$, and $ \avg{\comm{\xop(t)}{ \pop(t)}_{+}} = 2 \avg{\pop^2(0)} t/m$. We remark that the extra diffusive term $\sim t^{3}$ found in \eqcite{eq:moments} for the position fluctuations is a clear signature of a random force without damping, which is the case of the position-localization decoherence. It is also interesting to note that Eqs.~\eqref{eq:moments} depend only on $\Lambda$ and therefore could have been obtained with the simpler master equation \eqcite{eq:localization} which is, however, only valid in the short-distance limit. 

The position correlation function $\mathcal{C}(x,t) $ can be now computed using \eqcite{eq:solution} and \eqcite{eq:moments}. In particular, by taking into account that $ \hat \rho_{s}(t)$ is a Gaussian state, one can perform the integration over $y$ in \eqcite{eq:solution} and obtain
\be
\begin{split}
\mathcal{C}&(x,t) = \int_{-\infty}^{\infty}  \frac{dp }{2 \pi }  \mathcal{F}(p,x,t) \\
& \times \exp \left[- \frac{\avg{\pop^{2}}_{s}x^{2}+\avg{\xop^{2}}_{s}p^{2} - \avg{\comm{\xop}{\pop}_{+}}_{s} xp}{2 \hbar^{2}} \right].
\end{split}
\ee
Note that we do not explicitly write the time dependence of the moments in order to ease the notation. A simpler formula of $\mathcal{C}(x,t)$ for the short(long)-distance limit can be derived by using the corresponding approximation in $\mathcal{F}(p,x,t)$. This leads to
\be \label{Eq:coherence}
\frac{\mathcal{C}(x,t)}{\mathcal{C}(0,t)} \approx \left\{ \begin{array}{ll}
                  \exp[- x^{2}/\xi^{2}(t)]
, & x \ll 2a ,\\
                   \exp[- x^{2}/\xi^{2}_{s}(t) - \gamma t] 
, & x \gg 2 a. \end{array} \right.
\ee
We have defined the coherence lengths
\be \label{eq:coherencelength}
\xi^{2}(t)= \frac{8 \hbar^{2} \avg{\xop^{2}(t)}}{4 \avg{\xop^{2}(t)} \avg{\pop^{2}(t)} - \avg{ \comm{\xop(t)}{ \pop(t)}_{+}}^{2}},
\ee
and 
\be \label{eq:xis}
\xi^{2}_{s}(t) = \frac{8 \sigma^{2}(t)}{2 \bar n +1}.
\ee
$\xi_{s}$ is obtained by evaluating \eqcite{eq:coherencelength} with the unitary evolution given by the Schr\"odinger equation alone. Recall the definition of $\sigma^{2}(t)=x^{2}_{0} (1+ t^{2} \omega^{2})$ and that for an initial thermal state, $\avg{\xop^{2}(0)}=(2 \bar n+1)x^{2}_{0}$. 
\begin{figure}
\begin{center}
\includegraphics[width=\linewidth]{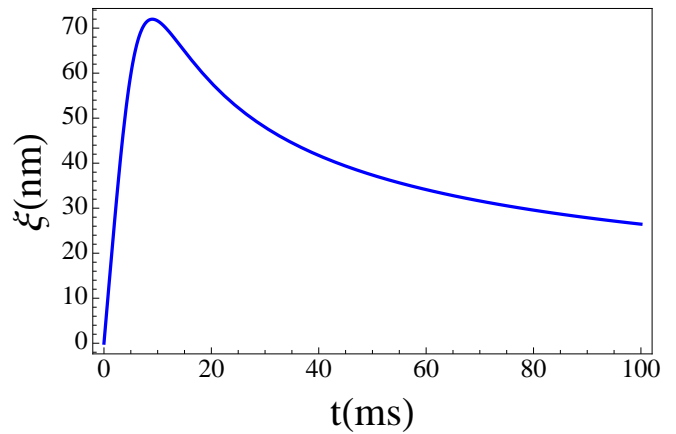}
\caption{(color online) As an example, the coherence distance $\xi(t)$, see \eqcite{eq:coherencelength}, is plotted for a sphere of $R=50$ nm at a bulk temperature of $200$ K taking into account the decoherence given by blackbody radiation, see Sec.~\ref{sec:bb}. Other experimental parameters are taken from Table~\ref{t:parameters}.}
\label{Fig:xi}
\end{center}
\end{figure}
While $\xi_{s}(t)$ increases monotonically in time, $\xi(t)$ has a maximum at 
\be \label{eq:tmax}
t_{\text{max}}= \left[  \frac{3 m (2 \bar n +1)  }{ 2 \Lambda \hbar \omega} \right]^{1/3},
\ee
which yields
\be \label{eq:ximax}
\xi_{\text{max}}= \sqrt{2} \left[\frac{2  \hbar \omega}{3 m \Lambda^{2}  (2 \bar n +1)  } \right]^{1/6}.
\ee
See Fig.~\ref{Fig:xi} for a particular example. Notice that the maximum coherence distance, according to \eqcite{Eq:coherence},   depends crucially on the saturation distance $a$ of the position-localization decoherence. 

As mentioned before, the coherence distance imposes some conditions on \acron. In particular, the superposition size $d$ has to be smaller than the coherence distance, namely, one requires $\mathcal{C}(d,t_{1})/\mathcal{C}(0,t_{1}) \sim 1$ in order to prepare a coherent superposition instead of a statistical mixture. This gives rise to conditions (v) and (vi) in Table~\ref{t:summary} depending on the ratio $d/(2 a)$.

\subsubsection{Visibility of the interference pattern}

The position-localization decoherence can also compromise the visibility of the interference pattern in the step \ref{step5} and \ref{step6} of \acron.  Using \eqcite{eq:solution}, one obtains that the time evolution of the position distribution $P(x,t)\equiv \bra{x} \hat \rho(t) \ket{x}$  is given by
\be \label{eq:blurring}
\begin{split}
P(x,t)&= \frac{1}{2 \pi \hbar}  \int_{-\infty}^{\infty} dp dx'    e^{\frac{\im p x}{\hbar}} {\mathcal{F}}(p) e^{-\frac{\im p x'}{\hbar}} P_{s}(x',t) \\
&= \frac{1 }{\sqrt{2 \pi \hbar}}  \int_{-\infty}^{\infty} dp  e^{\frac{\im p x}{\hbar}} {\mathcal{F}}(p)   \tilde P_{s}(p,t),
\end{split}
\ee
where $\tilde P_{s}(p,t) = \int dx e^{-\im p x'/\hbar} P_{s}(x,t) /\sqrt{2 \pi \hbar}$.
The position distribution without decoherence $P_{s}(x,t)$ oscillates with a wavelength given by $x_{f}=2 \pi \hbar t/(m d)$, which corresponds to the distances between the interference maxima. Thus, $\tilde P_{s}(p,t)$ has peaks at $p_{f}= \pm 2 \pi \hbar /x_{f}=m d/t$. Hence, the reduction of the interference peaks, which we use as a figure of merit for the visibility of the interference pattern, is given by
$ \mathcal{V} (t) \equiv  {\mathcal{F}}\left(2 \pi \hbar /x_{f} \right)= \exp \left[ - t \Theta \right]$,
where
\be \label{eq:theta}
\Theta=\gamma  - \gamma  \frac{ \sqrt{\pi} a}{ d }  \text{erf} \left[ \frac{  d }{ 2a} \right].
\ee
Note that $\Theta  \approx \Lambda d^{2}/3 $ in the limit  $d \ll 2a$ and $\Theta  \approx \gamma  $ in the limit  $d \gg 2a$. Therefore, the requirement $\Theta t_{2} \ll 1$ establishes the conditions (vii) and (viii) in Table~\ref{t:summary} depending on the ratio $d/(2a)$.

\subsection{Unavoidable sources of standard decoherence}

Let us now focus on the unavoidable decoherence given by scattering of air molecules and blackbody radiation. Decoherence due to environmental scattering is a well studied topic triggered by the work of Joos and Zeh~\cite{Joos1985}. For an extensive review on these topics, we refer the reader to the textbooks~\cite{Joos2003,Schlosshauer2007}. Here, we review the results needed for our analysis. Localization due to environmental scattering is described by a master equation of the type
\be \label{eq:MEexact}
\bra{\xx}\dot{ {\rho}}(t) \ket{\xx'} = \frac{\im}{\hbar} \bra{\xx} \comm{\hat \rho}{\hat H} \ket{\xx'} -F(\xx-\xx') \hat \rho (\xx,\xx'), 
\ee
where the decoherence function $F(\xx-\xx')$ depends on the distance $|\xx-\xx'|$ and can be expressed as~\cite{Schlosshauer2007}
\be
\begin{split}
F(\xx) =&  \int_{0}^{\infty} d q  \rho(q) v(q) \int \frac{d \nn d \nn'}{4 \pi}  \\
&\times \left( 1 - e^{\im q (\nn-\nn') \cdot \xx /\hbar}\right) |f(q \nn, q \nn')|^{2}.
\end{split}
\ee
The derivation assumes an infinitely massive object and the fact that the incoming particles are isotropically distributed in space. Here, $\rho(q)$ denotes the number density of incoming particles with magnitude of momentum equal to $q$, $v(q)=q/m_{a}$ ($v(q)=c$)  is the velocity of massive (massless) particles,  $|\nn|=|\nn'|=1$, and $f(q \nn, q \nn')$ is the elastic scattering amplitude. For further details, see Chapter 3 of~\cite{Schlosshauer2007}. The behavior of the function is very different depending on the ratio between $|\xx-\xx'|$ and the thermal wavelength of the scattering particles $\lambda_{\text{th}}$. In the long wavelength limit, $\lambda_{\text{th}}  \gg |\xx-\xx'|$, $F(\xx-\xx') \sim \Lambda |\xx-\xx'|^{2}$, whereas in the short wavelength limit, $\lambda_{\text{th}}  \ll |\xx-\xx'|$, one obtains the saturation of $F(\xx-\xx') \sim \gamma $. That is, above some critical distance each scattering event resolves the separation $|\xx-\xx'|$ and thereby provides which path information. This allows us to relate qualitatively and quantitatively the master equation \eqref{eq:MEexact} with the simpler one \eqref{eq:localizationgeneral} discussed in the previous subsections. This connection, which has been discussed previously in~\cite{Gallis1990,Vacchini2007}, is given by the following relations
\be \label{eq:connection}
a=\lambda_{\text{th}}/2 \hspace{1em} \text{and}  \hspace{1em} \gamma=  \lambda_{\text{th}}^{2} \Lambda.
\ee

In the analysis of decoherence due to environmental scattering one typically employs the long wavelength limit since it always  provides upper bounds on decoherence rates, even when one is in the short wavelength limit. This was the case, for instance, in the analysis performed in the optomechanical double slit proposal in~\cite{Romero-Isart2011c}. As shown below, the upper bounds for the case of scattering of air molecules were too loose, since one is in the saturation regime. This yielded the requirement of very low pressures.  The analysis performed in the following takes into account the saturation effect and yields much more feasible vacuum conditions. 

\subsubsection{Air molecules} \label{sec:gas}

\begin{figure}
\begin{center}
\includegraphics[width=\linewidth]{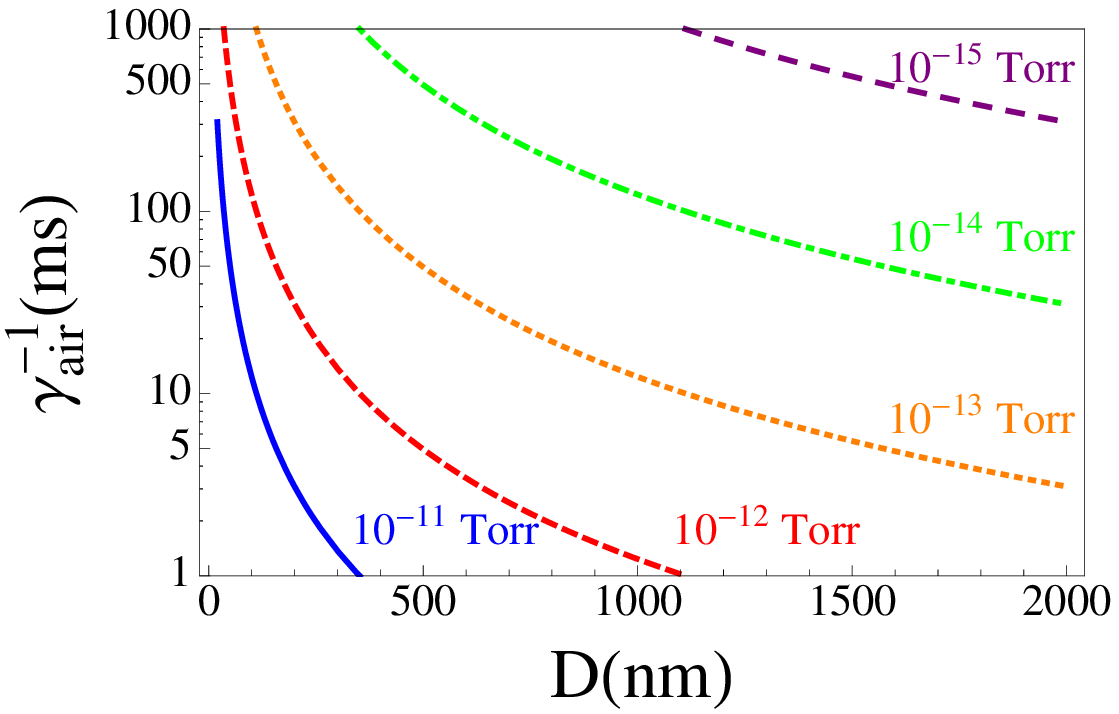}
\includegraphics[width=\linewidth]{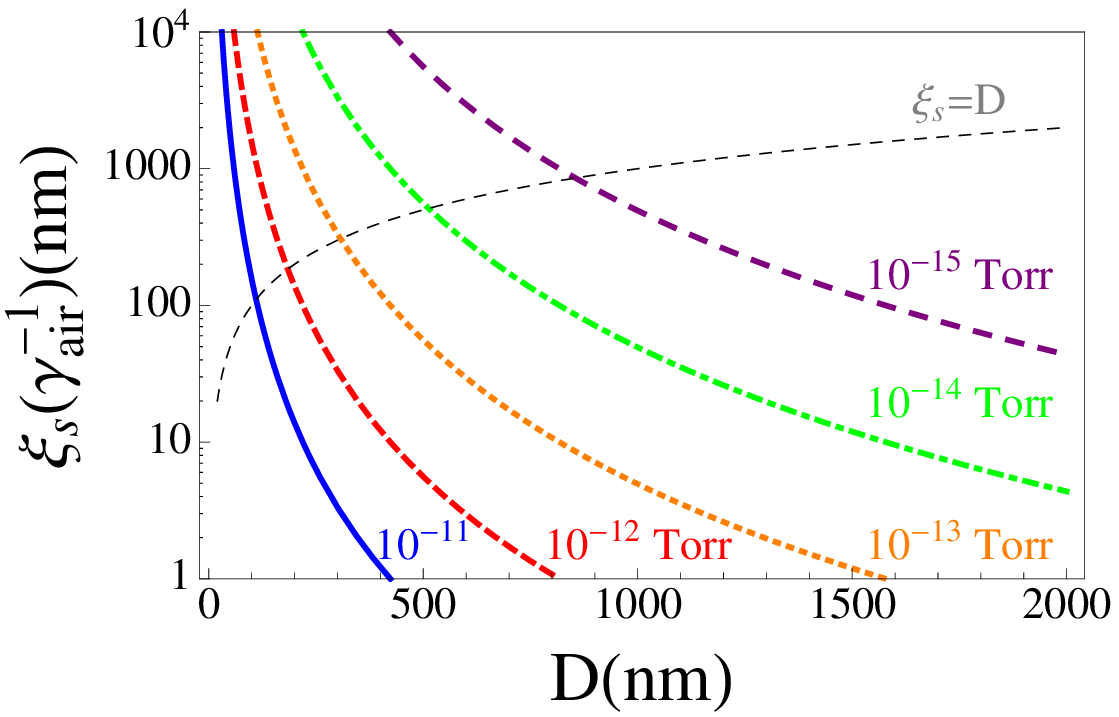}
\caption{(color online) Coherence time $1/\gamma_{\text{air}}$ (upper panel) and the corresponding coherence distance $\xi_{\text{s}}(1/\gamma_{\text{air}})$ (lower panel) taking into account the decoherence of air molecules as a function of the sphere's diameter and for environmental pressures of $P=10^{-11}$ Torr (solid blue), $10^{-12}$ Torr (dashed red), $10^{-13}$ Torr (dotted orange), $10^{-14}$ Torr (dotdashed green), and $10^{-15}$ Torr (largely dashed purple). Other experimental parameters are taken from Table~\ref{t:parameters}. In the lower panel, the thinner dashed gray line corresponds to the line  $\xi_{\text{s}}(1/\gamma_{\text{air}}) =D$. }
\label{Fig:Figair}
\end{center}
\end{figure}

The thermal wavelength of a typical air molecule, which is assumed to be in thermal equilibrium with an environment at temperature  $T$, is given by
$\lambda_\text{th}^{\text{air}}= 2 \pi \hbar/\sqrt{2 \pi m_{a}k_{b} T_{e}} \equiv 2 a_{\text{air}}$,
where $m_{a}$ is its mass. Using $m_{a} \sim 28.97$ amu and $T_{e}\sim4.5$ K, one obtains $2 a_{\text{air}}\sim 0.15$ nm. The localization parameter associated with scattering of air molecules in the long wavelength limit is given by~\cite{Schlosshauer2007}
\be \label{eq:locair}
\Delta_\text{air}= \frac{8 \sqrt{2 \pi} m_a \bar v  P R^2}{3 \sqrt{3}\hbar^2},
\ee
where $\bar v$ is the mean velocity of the air molecules, $P$ the pressure of the gas, and $R$ the radius of the sphere. Thus, using Eqs.~\eqref{eq:connection}, \eqref{eq:locair}, and the expression of $\lambda_\text{th}^{\text{air}}$, one obtains
\be
\gamma_{\text{air}}=    \frac{16 \pi \sqrt{2 \pi} }{ \sqrt{3}} \frac{  P  R^2}{ \bar v m_{a}}.
\ee
In the following, we will consider superpositions which are, at least, in the nanometer scale. Therefore,  we will use the short wavelength limit $d \gg 2a$ to account for the decoherence effect of air molecules.  The effect of this decoherence is shown in Fig.~\ref{Fig:Figair}, where the coherence time $\gamma^{-1}_{\text{air}}$ and the corresponding coherence distance $\xi_{s}(\gamma^{-1}_{\text{air}})$ are plotted as a function of the diameter of the sphere and for different pressures. Note that these quantities define the conditions (vi) and (viii) in Table~\ref{t:summary}. In particular, for sufficiently low pressures, large superpositions of the order of the size of the object are permitted. We remark again that in~\cite{Romero-Isart2011c} the saturation effect was not taken into account, and this gave rise to pressures of $10^{-16}$ Torr for spheres of $40$ nm, which turns out to be a very loose upper bound when taking into account the saturation effect. We will come back to this point in the optomechanical implementation of \acron~in Sec.\ref{sec:OMdoubleslit}.

\subsubsection{Blackbody radiation} \label{sec:bb}

The thermal wavelength for massless particles is given by
$\lambda_\text{th}^{\text{bb}} =  \pi^{2/3} \hbar c /(k_{b} T_{e}) \equiv 2 a_{\text{bb}}$,
which at temperatures $T\sim 4.5$ K takes the value of $\lambda_\text{th}^{\text{bb}} \sim 1$~mm. In this case the long wavelength limit can be employed since the superpositions considered will be always smaller than $\lambda_\text{th}^{\text{bb}}$. Recall that in this limit the relevant quantity is the localization parameter. This parameter has three contributions given by scattering, emission, and absorption of thermal photons, namely,  $\Lambda_\text{bb} = \Lambda_\text{bb,sc} + \Lambda_\text{bb,e} +\Lambda_\text{bb,a} $, which are given by
\be
\Lambda_\text{bb,sc}= \frac{8! \times 8  \zeta(9) c R^6}{9 \pi}  \left[ \frac{k_b T_e}{\hbar c} \right]^9 \text{Re} \left [\frac{\epsilon_\text{bb}-1}{\epsilon_\text{bb}+2} \right]^2,
\ee
and
\be
\Lambda_\text{bb,e(a)}= \frac{16 \pi^5 c R^3}{189} \left[ \frac{k_b T_{i(e)}}{\hbar c}\right]^6 \text{Im} \left [\frac{\epsilon_\text{bb}-1}{\epsilon_\text{bb}+2} \right].
\ee
We refer the reader to~\cite{Schlosshauer2007,Chang2010} for further details. Here,  $\zeta(x)$ is the zeta Riemmann function, $\epsilon_\text{bb}$ is the average dielectric constant, which is assumed to be time independent and relatively constant across the relevant blackbody spectrum, and 
 $T_{i}$ is the bulk temperature of the object, which might differ from $T_{e}$.

\begin{figure}
\begin{center}
\includegraphics[width=\linewidth]{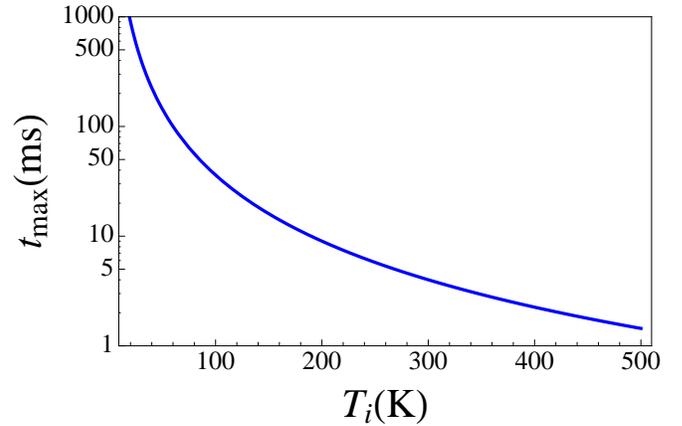}
\caption{(color online) $t_{\text{max}}$ taking into account blackbody radiation is plotted as a function of the internal temperature of the sphere. We used a sphere of radius of $50$ nm although the dependence on the radius is negligible. Other experimental parameters are taken from Table~\ref{t:parameters}.}
\label{Fig:BB}
\end{center}
\end{figure}

From the three contributions, the emission localization parameter is usually the dominant one since the internal temperature is usually larger than the external one, for instance, due to laser absorption during the optical manipulation of the sphere. In Fig.~\ref{Fig:BB} we plot the optimal time $t_{\text{max}}$, see \eqcite{eq:tmax} as a function of the internal temperature of the object. The dependence of $t_{\text{max}}$ on the size of the sphere is negligible. Comparing Fig.~\ref{Fig:BB} with the upper panel of Fig.~\ref{Fig:Figair}, one concludes that, for low pressure, the decoherence due to blackbody radiation will be dominant, specially, when the internal temperature is different from the external one which is supposed to be cryogenic (a few Kelvins).

\subsubsection{Limitations on the superposition size} 

\begin{figure}
\begin{center}
\includegraphics[width=\linewidth]{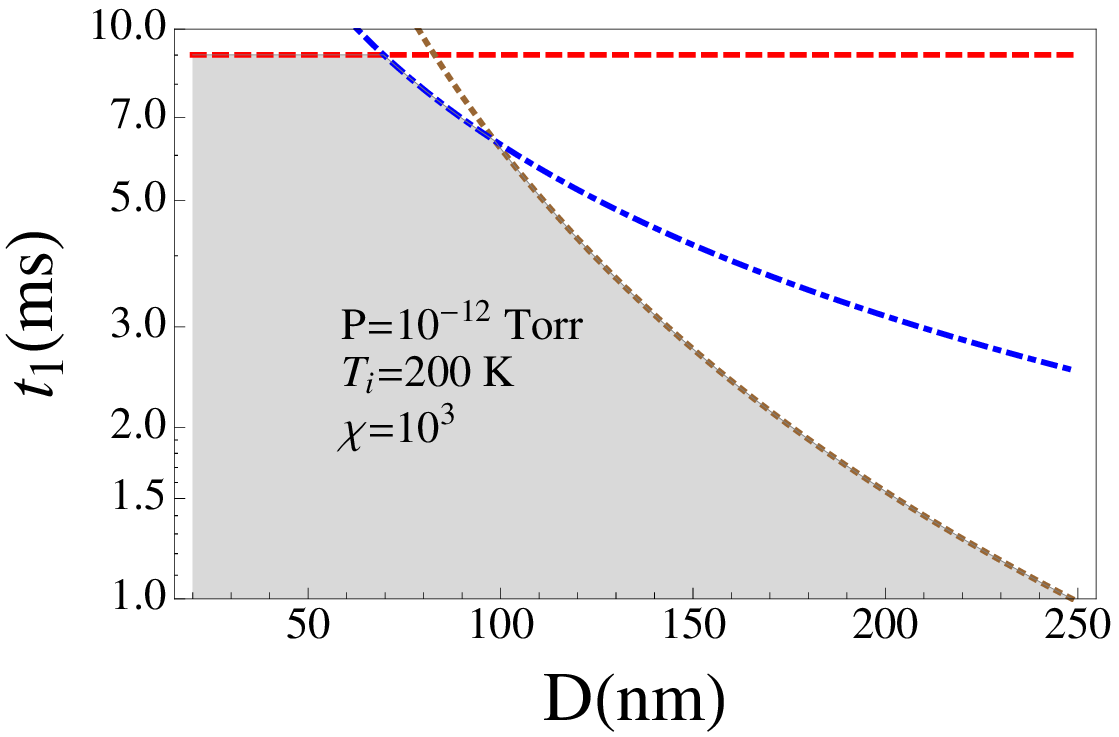}
\includegraphics[width=\linewidth]{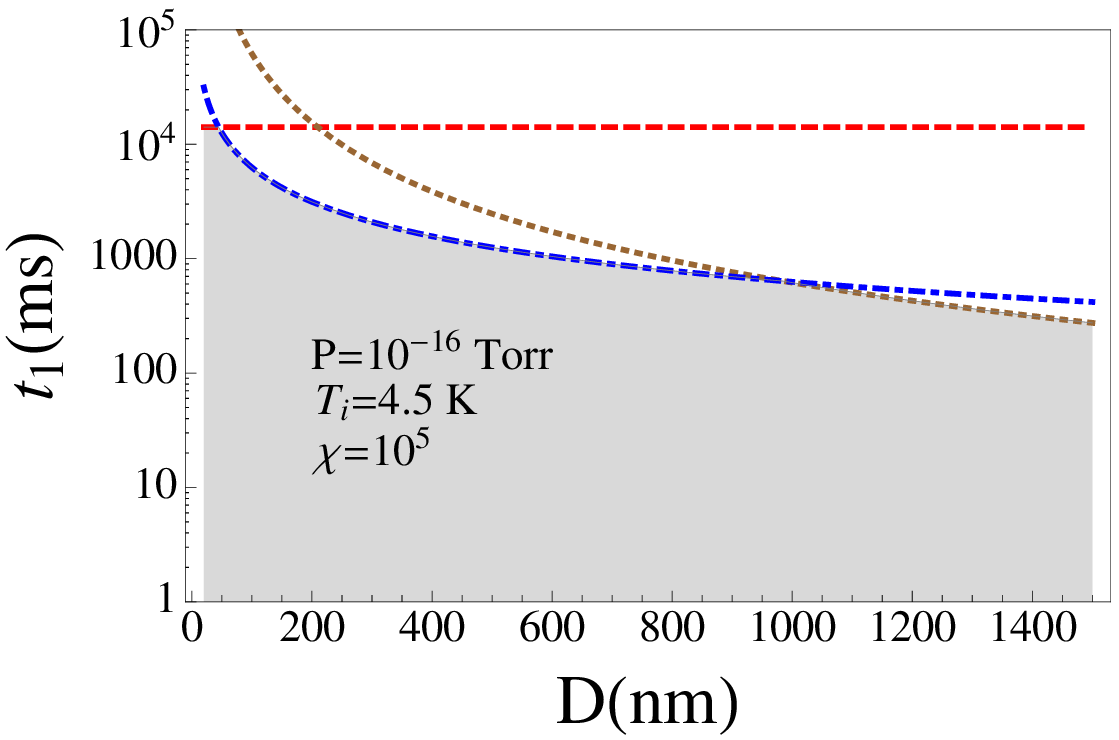}
\caption{(color online) Upper bounds for $t_{1}$ due to blackbody radiation and scattering of air molecules as a function of the diameter of the sphere for $P=10^{-12}$ Torr, $T_{i}=200$ K, and $\chi=1000$ ($P=10^{-16}$ Torr, $T_{i}=4.5$ K, and $\chi=10^{6}$) in the upper (lower) panel. $t_{\text{max}}$ (dashed red), see \eqcite{eq:tmax}, is condition (v) in Table~\ref{t:summary}, $0.05/\gamma_{\text{air}}$ (dotted brown) is condition (vi), and $\sqrt{2 t_{2} \chi/\omega}$ (dotdashed blue) is condition (iii). We used $t_{2}=0.1/\gamma_{\text{air}}$ and the others parameters given in Table~\ref{t:parameters}. The shadowed region corresponds to $t_{1}$ fulfilling all three conditions.}
\label{Fig:t1max}
\end{center}
\end{figure}

\begin{figure}
\begin{center}
\includegraphics[width=\linewidth]{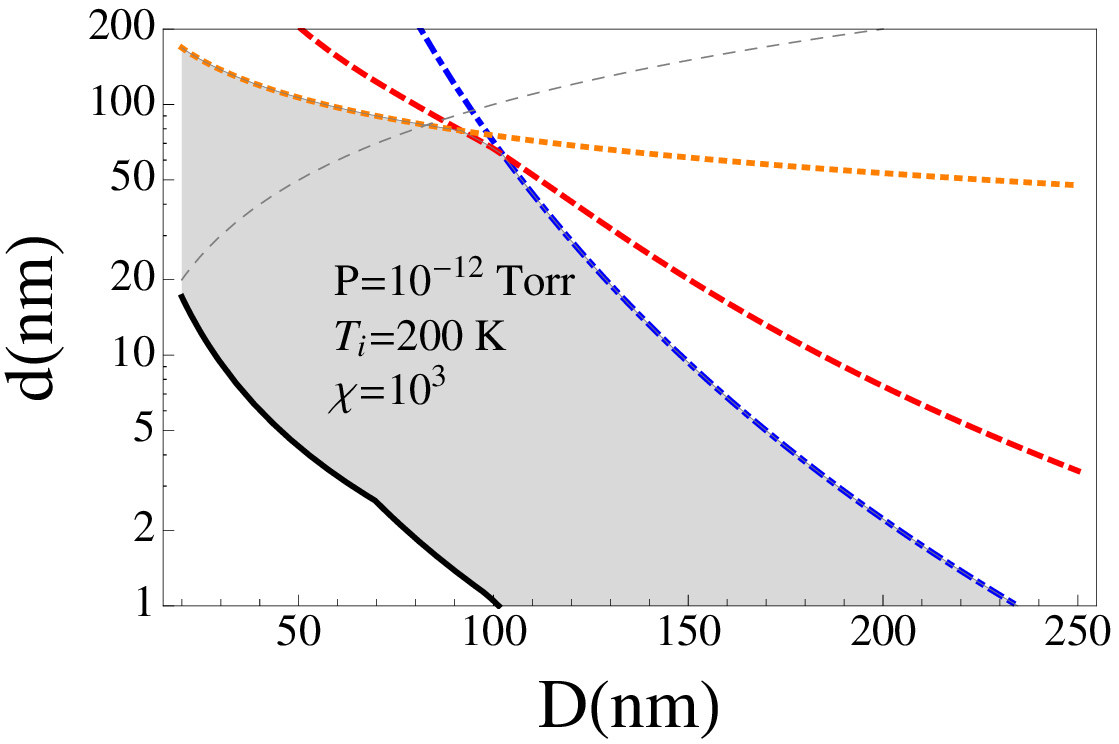}
\includegraphics[width=\linewidth]{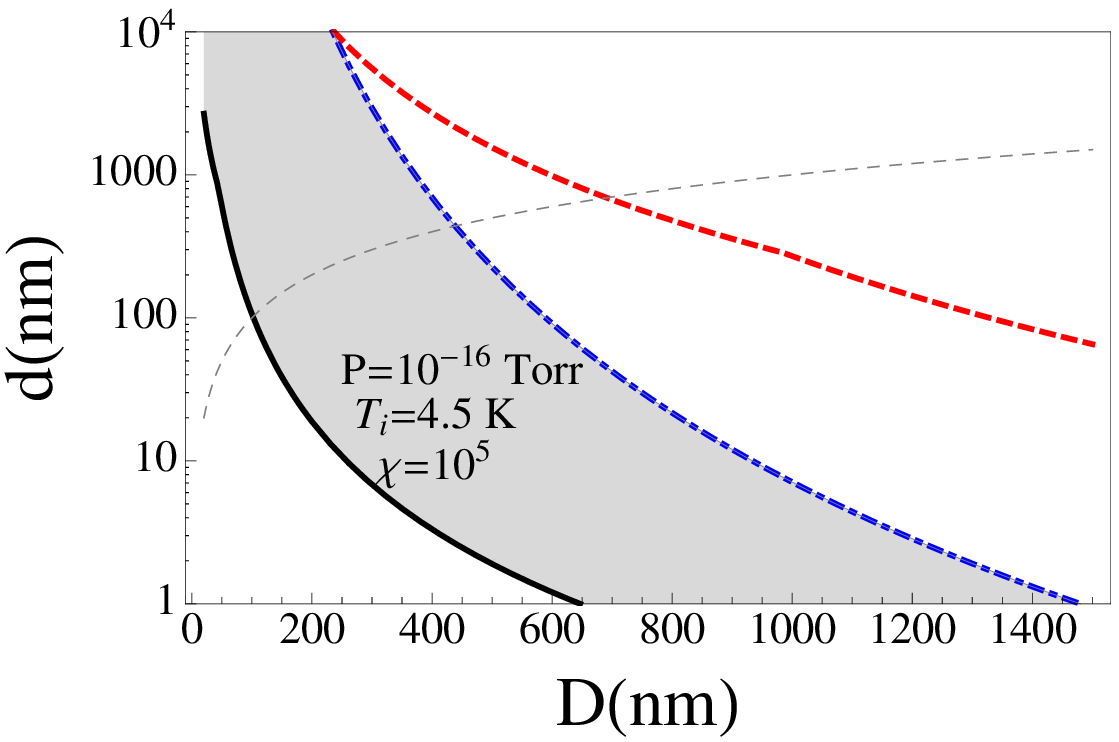}
\caption{(color online) Superposition distance $d$ as a function of the diameter $D=2R$ of the sphere  taking into account blackbody radiation and scattering of air molecules with $P=10^{-11}$ Torr, $T_{i}=200$ K, and $\chi=1000$ ($P=10^{-16}$ Torr, $T_{i}=4.5$ K, and $\chi=10^{6}$) in the upper (lower) panel and the others parameters given in Table~\ref{t:parameters}. Note the different scales for the upper and lower panels. According to Table~\ref{t:summary}, we plotted condition (ii) $ d > \sigma /\sqrt{\chi}$ (solid black), condition (iv) $d <  2 \pi \hbar t_{2}/(m \delta x)$ (dotdashed blue), condition (v) $ d < \xi (t_{1})$  (dashed red), and condition (vii) $d < \sqrt{3/(\Lambda t_{2})}$ (dotted orange). The shadowed region corresponds to $d$ fulfilling all four conditions and thin dashed gray line to $d=D$. We used $t_{2}=0.1/\gamma_{\text{air}}$ and $t_{1}= \min \left\{ \sqrt{2 t_{2} \chi /\omega}, t_{\text{max}}, 0.05/\gamma_{\text{air}} \right\}$.}
\label{Fig:Limitationsond}
\end{center}
\end{figure}

Let us now summarize the operational parameter regime of \acron~taking into account standard sources of decoherence. We compute the lower and upper bounds for $d$ for allowed values of  $t_{1}$ and $t_{2}$ according to Table~\ref{t:summary}. We will use two sets of experimental parameters for the pressure, the internal temperature, the measurement strength, and the common ones given in Table~\ref{t:parameters}. The first set, which is assumed to be feasible, assumes an environmental pressure of $P=10^{-12}$ Torr, internal temperature of the object $T_{i}=200$ K, and measurement strength $\chi=1000$, whereas the challenging set assumes $P=10^{-16}$ Torr, $T_{i}=4.5$ K, and $\chi=10^{6}$.

Figure~\ref{Fig:t1max} shows the upper bounds for $t_{1}$ given by conditions (iii), (v), and (vi) in Table~\ref{t:summary}, considering $t_{2}=0.1/\gamma_{\text{air}}$. For the feasible set, $t_{1}$ in the few milliseconds are possible for spheres up to a diameter of $250$ nm. For the challenging set, much larger timescales of the order of seconds are possible for objects in the micrometer regime. In both cases, pressure is the limiting factor for the internal temperatures considered. Note that, in both cases, these values correspond to very long coherence times comparing to typical quantum-mechanical experiments.

In Fig.~\ref{Fig:Limitationsond} we plot the superposition size that fulfills conditions (ii), (iv), (v), and (vii) in Table~\ref{t:summary} for $t_{2}=0.1/\gamma_{\text{air}}$ and $t_{1}= \min \left\{ \sqrt{2 t_{2} \chi /\omega}, t_{\text{max}}, 0.05/\gamma_{\text{air}} \right\}$, which ensures the fulfillment of the restrictions imposed into $t_{1}$ and $t_{2}$. Hereafter, we will call this plot a ``$d$ vs. $D$'' diagram. For the feasible set, superpositions larger than the size of the sphere are in principle possible for objects of the order of $50$ nm with much lower pressures than those used in~\cite{Romero-Isart2011c}. For the challenging set, larger objects and larger superpositions are obtained (note the different scale in the plot). It is important to remark that in both cases the limitation is given by condition (iv), which reads $d <  2 \pi \hbar t_{2}/(m \delta x)$ and is related to the resolution of the interference fringes. This shows that while coherence times are very large, the dynamics resulting in the spreading of the wave function are very slow for large masses. This hints at possible improvements of  \acron~using more efficiently the long coherence times allowed by the unavoidable sources of decoherence considered here.

\section{Collapse models} \label{Sec:CollapseModels}

In this Section we discuss the possibility to use \acron~to test some of the most paradigmatic collapse models. Remark that any experimental evidence of these would imply a breakdown of the theory of quantum mechanics. In the following we consider the unavoidable sources of decoherence discussed in the previous Section and, one these grounds, we determine the experimental parameters required to falsify a given collapse model by the observation of the interference pattern. We note that the corroboration of the collapse model is more challenging than its falsification, since one must discard that standard decoherence, in any of its forms, is responsible for the disappearance of the interference pattern. 

We shall not review the extensive literature on collapse models; instead, we will focus on some of the most discussed ones in the literature, and for each of them, we will provide a brief summary of their prediction. In particular, we shall express them in a common form, namely as a master equation describing position-localization decoherence; this will allow us to apply the results of Sec.~\ref{sec:posloc}, and thereby to discuss the possibility to test them using \acron.

\subsection{Continuous spontaneous localization}

We start with the Continuous Spontaneous Localization (CSL) model~\cite{Pearle1989,Ghirardi1990a}, which is the best developed collapse model at present~\cite{Bassi2003}. This model builds upon the previous works of Ghirardi, Rimini, and Weber (GRW)~\cite{Ghirardi1986}, Pearle~\cite{Pearle1976}, Gisin~\cite{Gisin1984}, and it bears some similarity to the works of Gisin~\cite{Gisin1989} and Di\'osi~\cite{Diosi1988,Diosi1989}. The model is constructed by adding a non-linear stochastic term to the Schr\"odinger equation. This term predicts a localization whose strength is directly proportional to the mass of the object. At the same time, it is constrained by the fact that the equation must reproduce all phenomenology of quantum mechanics for small objects. This introduces two phenomenological constants that are bounded by experimental evidence.

More specifically, within the CSL model, the master equation describing the wavefunction of $N$ particles is given in the position representation $\ket{\xx_{1},\ldots,\xx_{N}}\equiv \ket{x}$ by~\cite{Pearle1989,Ghirardi1990a,Collett2003}
\be
\bra{x}\dot{ {\rho}}(t) \ket{x'} = \frac{\im}{\hbar} \bra{\xx} [\hat \rho,\hat H] \ket{x'}-\Gamma_{\text{CSL}}(x,x') \bra{\xx} \hat \rho(t) \ket{x'},
\ee
where
\be \label{eq:GammaCSL}
\begin{split}
\Gamma_{\text{CSL}}&(x,x') = - \frac{\gamma^{0}_{\text{CSL}}}{2} \sum_{i,j=1}^{N}  \frac{m_{i} m_{j}}{m^{2}_{0}} \times \\
&\times \left[ \Phi(\xx_{i}- \xx_{j}) + \Phi(\xx'_{i}- \xx'_{j}) -2 \Phi(\xx_{i}- \xx'_{j}) \right].
\end{split}
\ee
Here, $m_{0}$ is the mass of a nucleon, $\gamma^{0}_{\text{CSL}}$ is the single nucleon collapse rate, and 
\be
\Phi(\rr) = \exp \left[ - \frac{|\rr|^{2}}{4 a^{2}_{\text{CSL}}}\right]
\ee
is the localization function with $a_\text{CSL}$ being the localization distance. Note that for the single nucleon case, \eqcite{eq:GammaCSL} reads
\be
\Gamma_{\text{CSL}}(\xx,\xx') = \gamma^{0}_{\text{CSL}} \left(1 -\exp \left[ - \frac{|\xx-\xx'|^{2}}{4 a^{2}_{\text{CSL}}}\right]   \right)
\ee
which has the same form as \eqcite{eq:Gamma}. The parameters $\gamma^{0}_{\text{CSL}}$ and $a_{\text{CSL}}$ are the two phenomenological constants of the model. Their value is bounded by both experimental data and ``philosophical'' reasons;  see~\cite{Feldmann2011} for a recent discussion. The standard values originally proposed in the GRW model~\cite{Ghirardi1986} are $a_{\text{CSL}} =100$ nm  and $\gamma^{0}_{\text{CSL}} = 10^{-16}$ Hz. However, the value of $\gamma^{0}_{\text{CSL}}$ has been recently reconsidered by Adler and is predicted to be 8 to 10 orders of magnitude larger~\cite{Adler2007,Adler2009} than the original one of $10^{-16}$ Hz, a prediction not yet confronted by up-to-date experiments~\cite{Adler2009}. We will however consider here the original values for the sake of comparison. 

The decoherence factor of the CSL model, \eqcite{eq:GammaCSL} can be obtained for the center-of-mass of a solid sphere of mass $m$, volume $V$, and homogeneous mass density. We use the results given in Ref.~\cite{Collett2003} where an analysis of the CSL model for the free propagation of a solid mass is discussed. In this case, the decoherence factor (\cf~\eqcite{eq:GammaCSL}) takes the form
\be \label{eq:GammaCSLSphere}
\begin{split}
\Gamma_{\text{CSL}}&(\xx,\xx') = - \gamma^{0}_{\text{CSL}}  \frac{m^{2}}{m^{2}_{0}} \times \\
&\times \int_{V} \frac{d \rr d \rr'}{V^{2}} \left[ \Phi(\rr- \rr') - \Phi(\rr- \rr' +\xx -\xx') \right].
\end{split}
\ee
In order to approximate \eqcite{eq:GammaCSLSphere} to \eqref{eq:Gamma}, we can extract the localization parameter in the expression given for the free evolution of the position fluctuation, which is given by~\cite{Collett2003}
\be \label{eq:CSLmoments}
\avg{\xop^{2}(t)}=\avg{\xop^{2}(t)}_{s} + \frac{\gamma^{0}_\text{CSL} \hbar^2 f(R/a_{\text{CSL}}) t^{3}}{6 m_{0}^{2} a_{\text{CSL}}^{2}}.
\ee
By comparing it with \eqcite{eq:moments}, one obtains 
\be
\Lambda_{\text{CSL}}=\frac{m^{2}}{m_{0}^{2}} \frac{\gamma^{0}_\text{CSL}}{4  a^{2}_{\text{CSL}}}  f(R/a_{\text{CSL}}),
\ee
where the function $f(x)$ is given by 
\be
f(x)=  \frac{6}{x^{4}} \left[ 1- \frac{2}{x^{2}} + \left(1+\frac{2 }{x^{2}} \right) e^{-x^{2}}\right],
\ee
and has the following limits $f(x\rightarrow 0)=1$, $f(1)=0.62$, and $f(x\rightarrow \infty) = 6/x^{4}$. Thus, recalling that $\Lambda=\gamma/(4 a^{2})$, one obtains the collapse rate
\be \label{eq:gammaCSL}
\gamma_{\text{CSL}}=  \frac{m^{2}}{m_{0}^{2}} \gamma^{0}_\text{CSL} f(R/a_{\text{CSL}}).
\ee
Note that the rate $\gamma_{\text{CSL}}$ grows quadratically with the number of nucleons for spheres smaller than $2 a_{\text{CSL}}$. 

\begin{figure}
\begin{center}
\includegraphics[width=\linewidth]{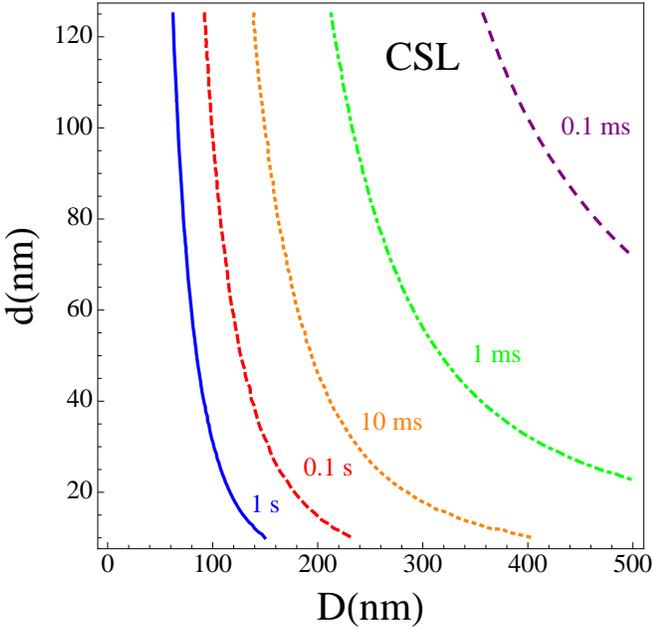}
\caption{(color online). Different values of the coherence time $1/ (\Lambda_{\text{CSL}} d^{2})$ as a function of the sphere's diameter $D$ and the superposition distance $d$. Other physical parameters, such as the density of the sphere, are taken from Table~\ref{t:parameters}. }
\label{Fig:PotentialCSL}
\end{center}
\end{figure}

To grasp the strength of the exotic decoherence by this model, we plot in in Fig.~\ref{Fig:PotentialCSL}   the value of the coherence time $1/ (\Lambda_{\text{CSL}} d^{2})$ of a superposition of size $d \ll 2 a_{\text{CSL}}$ as a function of the sphere's diameter $D$ and the superposition distance $d$. Coherence times of the order of milliseconds are obtained for objects of 300 nm and superpositions of tens of nanometers. Note that these coherence times would be strongly reduced by using the enhancement of the localization rate $\gamma^{0}_{\text{CSL}}$ predicted by Adler~\cite{Adler2007,Adler2009}.

\subsection{Quantum gravity}

Ellis, Mohanty, Mavromatos, and Nanopoulos suggested in~\cite{Ellis1989,Ellis1992} that quantum gravity (QG) can induce the collapse of the wavefunction of sufficiently massive objects. They argue that the collapse is induced by the interaction of the massive object with topologically non-trivial spacetime configurations (wormholes) which are small compared to physical scales but much larger than the Planck scale. From the point of view of quantum information theory, this decoherence mechanism can be understood as the result of the center-of-mass of the object becoming entangled with some degrees of freedom belonging to these wormholes which are unaccessible and therefore have to be traced out. Indeed, this provides a localization effect analogous to the one induced by the interaction with the environment. 

In this model, the quantum wormholes are assumed to be in a gaussian state in momentum space, and have zero mean momentum with a spread given by $\Delta \sim c m_{0}^{2}/(\hbar m_{P})\sim 10^{-3} $ m, where $m_{0}$ is the mass of a nucleon and $m_{P}$ is the Planck mass. For distances smaller than $1/\Delta$, the off-diagonal elements of the density matrix in position basis decay as $\bra {x} \hat \rho(t) \ket{x'} \propto \exp \left[ -\Lambda^{0}_{\text{QG}} (x-x')^{2} t\right]  \bra {x} \hat \rho(0) \ket{x'}$, where the localization parameter is given by
\be
\Lambda^{0}_{\text{QG}} = \frac{c^{4}}{\hbar^{3} }\frac{m^{6}_{0}}{m^{3}_{P}}. 
\ee
Note that the localization distance is very large since it is given by $a_{\text{QG}}=1/(2 \Delta) \sim 10^{3}$ m, hence,  the decoherence does not saturate within the typical wave function spreadings. The collapse rate for the single nucleon is thus given by $\gamma^{0}_{\text{QG}} = 4 a^{2}_{\text{QG}}\Lambda^{0}_{\text{QG}}$. It is remarkable that this model~\cite{Ellis1989,Ellis1992}, which is based on quantum gravity, converts the CSL model into a parameter-free model. The extension of the single nucleon case to the solid sphere can be obtained by retrieving the result given in \eqcite{eq:gammaCSL}. Based on this, the localization parameter of a solid sphere predicted by this model is given by
\be \label{eq:gammaQG}
\Lambda_{\text{QG}}= \frac{c^{4}}{\hbar^{3} }\frac{m^{2} m^{4}_{0}}{m^{3}_{P}},
\ee
where we have used $f(R/a_{\text{QG}})=1$ since $R \ll a_{\text{QG}}$ for the spheres considered here.

\begin{figure}
\begin{center}
\includegraphics[width=\linewidth]{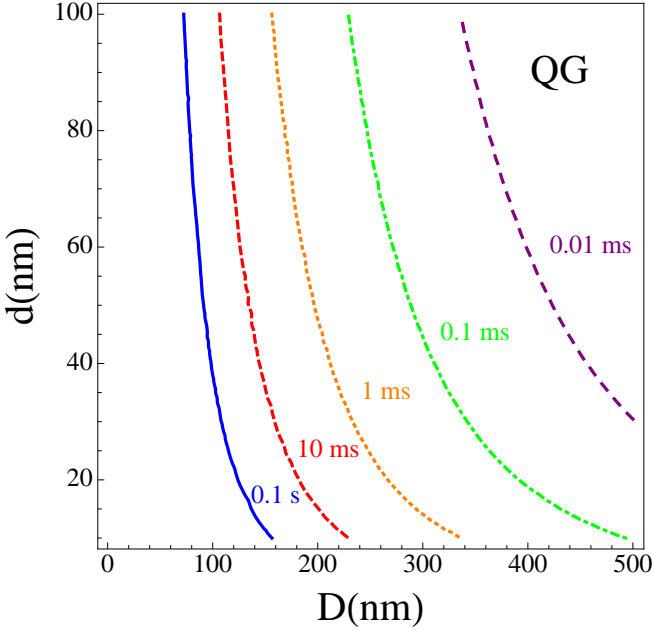}
\caption{(color online). Different values of the coherence time $1/ (\Lambda_{\text{QG}} d^{2})$ as a function of the sphere's diameter $D$ and the superposition distance $d$. Other physical parameters are taken from Table~\ref{t:parameters}.}
\label{Fig:PotentialQG}
\end{center}
\end{figure}

As previously, we plot in Fig.~\ref{Fig:PotentialQG} different values of the coherence time $1/ (\Lambda_{\text{QG}} d^{2})$ as a function of $D$ and $d$. By comparing these results with the ones given by the CSL model in Fig.~\ref{Fig:PotentialQG}, we notice that the decoherence effect is slightly stronger, but nonetheless very similar.

\subsection{von Neumann-Newton equation}

In the last 30 years, many authors have investigated the possible role of the Newtonian gravity in the collapse of the wavefunction. From these, the independent but similar works of Di\'osi and Penrose (DP) are the most famous ones~\cite{Diosi1984,Diosi1987,Diosi1989, Moroz1998,Penrose1996,Diosi2005,Diosi2007}. The prediction of the model can be casted into the so-called von Neumann-Newton equation, which can be expressed as~\cite{Diosi1987,Diosi1989,Diosi2007}
\be
\bra{\xx}\dot{ {\rho}}(t) \ket{\xx'} = \frac{\im}{\hbar} \bra{\xx} [\hat \rho,\hat H] \ket{\xx'}-\Gamma_{\text{DP}}(\xx,\xx') \bra{\xx} \hat \rho(t) \ket{\xx'},
\ee
where 
\be \label{eq:GammaDP}
\Gamma_{\text{DP}}(\xx,\xx') = \frac{U_{g}(\xx,\xx) - U_{g}(\xx',\xx')+2U_{g}(\xx,\xx')}{2 \hbar}.
\ee
Here $U_{g}(\xx,\xx')$ is the Newtonian interaction between two mass densities corresponding to two spheres centered at position $\xx$ and $\xx'$, respectively. This reads
\be
U_{g}(\xx,\xx') = - G \int \frac{f(\rr |\xx) f(\rr' | \xx')}{|\rr - \rr'|} d \rr d \rr',
\ee
where $f(\rr| \xx)$ is the mass density at location $\rr$ for the sphere centered at $\xx$. For a rigid homogeneous ball, the mass density is uniform and equals $\bar f= 3 M/(4 \pi R^{3})$. In this case, the decoherence function of \eqcite{eq:GammaDP} depends on the relative distance $|\xx-\xx'|$ and presents the following limits~\cite{Diosi1987}
\be \label{eq:ak}
\Gamma_{\text{DP}}(x) = \left\{ \begin{array}{ll}
                  G m^{2} /(2 R^{3} \hbar) x^{2}
, & x \ll  R ,\\
              6 G m^{2} / (5R \hbar)
, & x \gg  R. \end{array} \right.
\ee
The quadratic dependence at short distances allows us to identify the localization parameter of the model
\be
\Lambda_{\text{DP}}= \frac{G m^{2} }{2 R^{3} \hbar },
\ee
as well as the saturation distance at $2 a_{\text{DP}}=R$.

\begin{figure}
\begin{center}
\includegraphics[width=\linewidth]{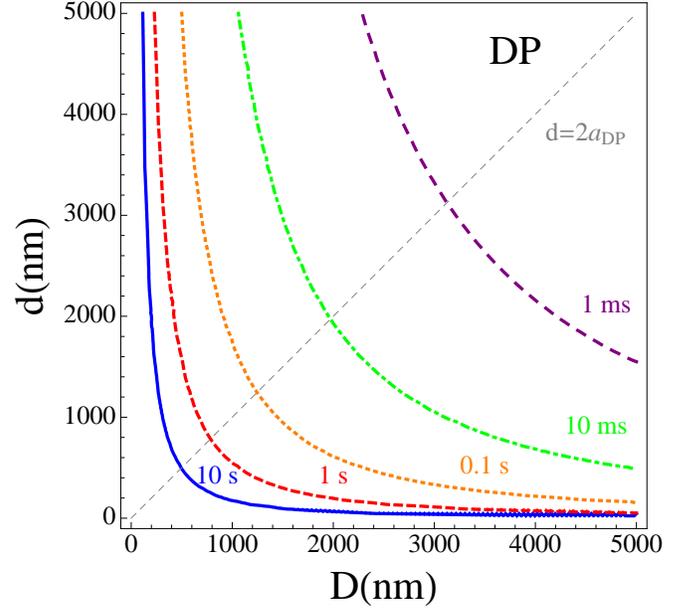}
\caption{(color online) Different values of the coherence time $1/ (\Lambda_{\text{DP}} d^{2})$ as a function of the sphere's diameter $D$ and the superposition distance $d$.  The dashed gray line marks the limit where the coherence time is not longer valid since, for $d > 2 a_{\text{DP}}$, it would then be given by $1/\gamma_{\text{DP}}$. Other physical parameters are taken from Table~\ref{t:parameters}.}
\label{Fig:PotentialDP}
\end{center}
\end{figure}

The strength of this model is much weaker than the CSL and the QG, see Fig.~\ref{Fig:PotentialDP} where $1/ (\Lambda_{\text{DP}} d^{2})$ is plotted as a function of $D$ and $d$. Coherence times of the order of milliseconds are only obtained for objects of few microns prepared in superpositions within the micrometer scale.

Let us point out that the decoherence rate given by this collapse model can be strongly enhanced by considering the mass density at the microscopic level~\cite{Diosi2007}. For instance, by modeling the fine structure beyond the constant average mass as a conglomerate of identical small balls of mass $m_{0}$ and radius $r_{0}$, the parameters of the model are given by~\cite{Diosi2007} $2 \tilde a_{\text{DP}}=r_{0}$ and
\be
\tilde \Lambda_{\text{DP}}= \left( \frac{R}{r_{0}}\right)^{3} \Lambda_{\text{DP}}.
\ee
Thus, since $r_{0}$ is typically chosen much smaller than $R$, the localization parameter is greatly enhanced. This is used, for instance, in the Marshall proposal to test the Penrose model using the small delocalization of a micro-mirror~\cite{Marshall2003,Kleckner2008}. However, this choice is controversial since it does not only convert the parameter free model into a one parameter ($r_{0}$) model, but the von-Neumann-Newton equation also becomes divergent for point-like particles~\cite{Ghirardi1990,Diosi2005,Diosi2007}. This yields unphysical results such as the non-conservation of energy, specially when $r_{0} \ll 100$ nm (see \cite{Ghirardi1990,Diosi2007}). For this reason, while it is not clear how to choose the mass distribution, in this article we assume the well-behaved case of a solid homogeneous density; this comes at the price of making formidable the possibility of falsifying the model,  as shown below.

\subsection{Imprecise space-time}

Finally, we also consider the K-model, named after K\'arolyh\'azy, who  introduced one of the first collapse models already the 1960's~\cite{Karolyhazy1966,Karolyhazy1974}. The model builds upon the insight that the sharply determined structure of space-time is incompatible with quantum mechanics and general relativity: according to quantum mechanics, the position and the velocity of an object cannot have deterministic values simultaneously, while general relativity states that the space-time structure is determined by the positions and velocities of the masses. 

We base our approach on the article of Frenkel~\cite{Frenkel1990} who provides a very clear review of the K-model and its relation to the CSL model. The prediction of the K-model in the so-called ``no-breathing limit''~\cite{Frenkel1990} is given by the following master equation
\be 
\dot \rho(t) = \frac{\im}{\hbar} \comm{\hat \rho(t)}{ \hat H} - \Lambda_{K} \comm{\xop}{\comm{\xop}{\hat \rho(t)}}.
\ee
This corresponds to a position-localization decoherence with a localization distance $2 a_{K} \rightarrow \infty$, a localization rate $\gamma_{K} \rightarrow 0$, and a localization parameter $\gamma_{K}/(4 a_{K}^{2} )\rightarrow \Lambda_{K}$. The localization parameter for a solid sphere of mass $m$ is given by~\cite{Frenkel1990}
\be
\Lambda_{K}=\frac{\hbar }{8 m a^{4}_{c} },
\ee
where
\be
 a_{c} = \left\{ \begin{array}{ll}
                 \left( \frac{R}{l_{P}}\right)^{3/2} l_{C}  
, & R > a_{K} ,\\
              \left( \frac{l_{C}}{ l_{P}}\right)^{2} l_{C}
, & R < a_{K}. \end{array} \right.
\ee
Here $l_{P}=\sqrt{G \hbar /c^{3}}$ is the Planck length and $l_{C}= \hbar/(mc)$ the Compton wavelength. Note that this model is also parameter-free.

\begin{figure}
\begin{center}
\includegraphics[width=\linewidth]{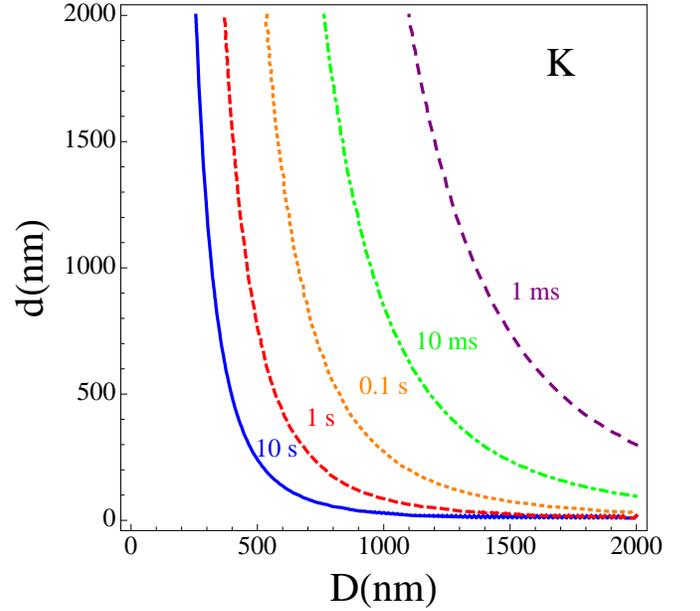}
\caption{(color online) Different values of the coherence time $1/ (\Lambda_{\text{K}} d^{2})$ as a function of the sphere's diameter $D$ and the superposition distance $d$. Other physical parameters are taken from Table~\ref{t:parameters}.}
\label{Fig:PotentialK}
\end{center}
\end{figure}

As shown in Fig.~\ref{Fig:PotentialK}, the strength of this model is also weaker than the CSL and the QG, and only slightly stronger than the DP.  Coherence times of the order of milliseconds are obtained for  spheres with diameter between one and two micrometers prepared in superpositions smaller than one micrometer. 

\subsection{Experimental test} \label{sec:experiments}

\begin{figure}
\begin{center}
\includegraphics[width=\linewidth]{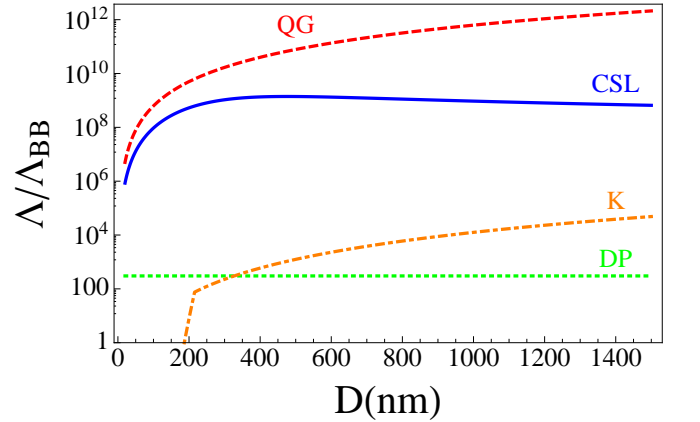}
\caption{(color online). Localization parameter $\Lambda$ for the different collapse models (CSL, QG, K, and DP) in units of the localization parameter provided by blackbody radiation, assuming a bulk temperature of $T_{i}=4.5$ K, as a function of the diameter of the sphere. Other parameters are taken from Table~\ref{t:parameters}.}
\label{Fig:CollapsevsBB}
\end{center}
\end{figure}

Let us now address the possibility to test these collapse models using \acron. The localization parameter and the localization distance of each model are summarized in Table~\ref{t:parametersCM}. %
\begin{table}
\begin{center}
\begin{ruledtabular}
\begin{tabular}{ | l | l | l | }
		&2a			&$\Lambda$ \\ \hline \hline
CSL  	&$\sim200$ nm		&  $ \gamma^{0}_\text{CSL} m^{2} f(R/a_{\text{CSL}}) /( 4  a^{2}_{\text{CSL}} m_{0}^{2})  $		\\ \hline
QG  		& $1/\Delta \sim 10^{3}$ m			& 	$c^{4}  m^{2} m^{4}_{0}/(\hbar^{3} m^{3}_{P})$	\\ \hline
DP	  	&R			&$  G m^{2}  /(2 R^{3} \hbar)$		\\ \hline
K	  	&$\infty$		& $\hbar /(8 m a^{4}_{c})$		\
\end{tabular}
\end{ruledtabular}
\caption{\label{t:parametersCM} Summary of the decoherence parameters predictied  by different collapse models. Recall that, for the CSL model we take the original value of $\gamma^{0}_{\text{CSL}} \approx 10^{-16}$ Hz. }
\end{center}
\vspace{-0.6cm}
\end{table}
For the sake of comparison, we plot the ratio between their localization parameter with the localization parameter provided by blackbody radiation in Fig.~\ref{Fig:CollapsevsBB}, where we have assumed a bulk temperature of $T_{i}=4.5$ K (recall Sec.~\ref{sec:bb}). While all collapse models provide a stronger localization rate than the blackbody radiation, the CSL and QG model are many orders of magnitude stronger than the DP and the K-model. Actually, the standard decoherence given by blackbody radiation is comparable to the one predicted by the DP and K-model for a bulk temperature of $20$ K. 
Since standard decoherence will limit the superpositions to be smaller than the localization distance, the localization parameter $\Lambda$ is the only relevant parameter.

\begin{figure}
\begin{center}
\includegraphics[width=\linewidth]{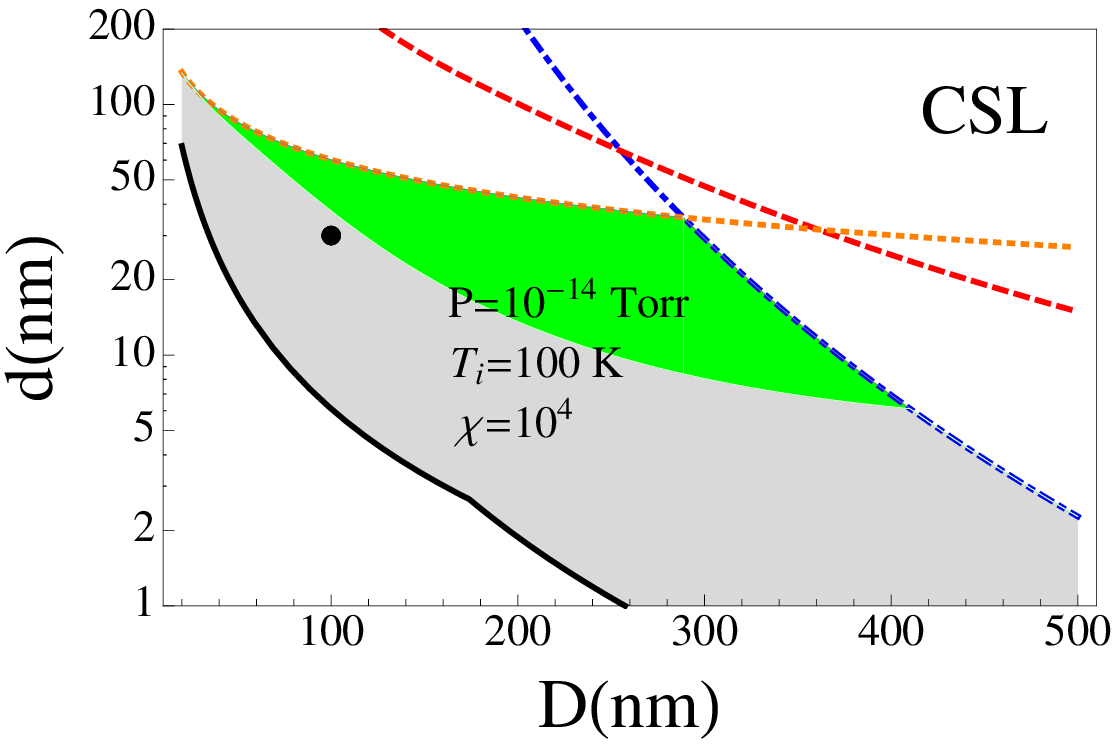}
\includegraphics[width=\linewidth]{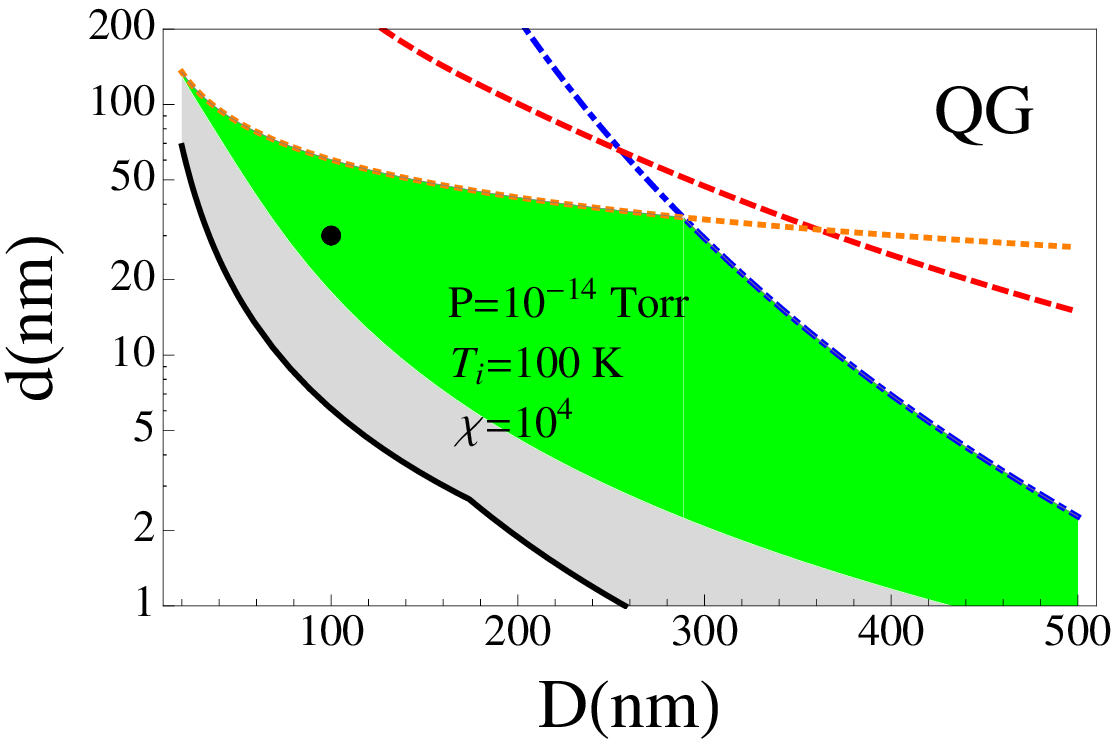}
\includegraphics[width=\linewidth]{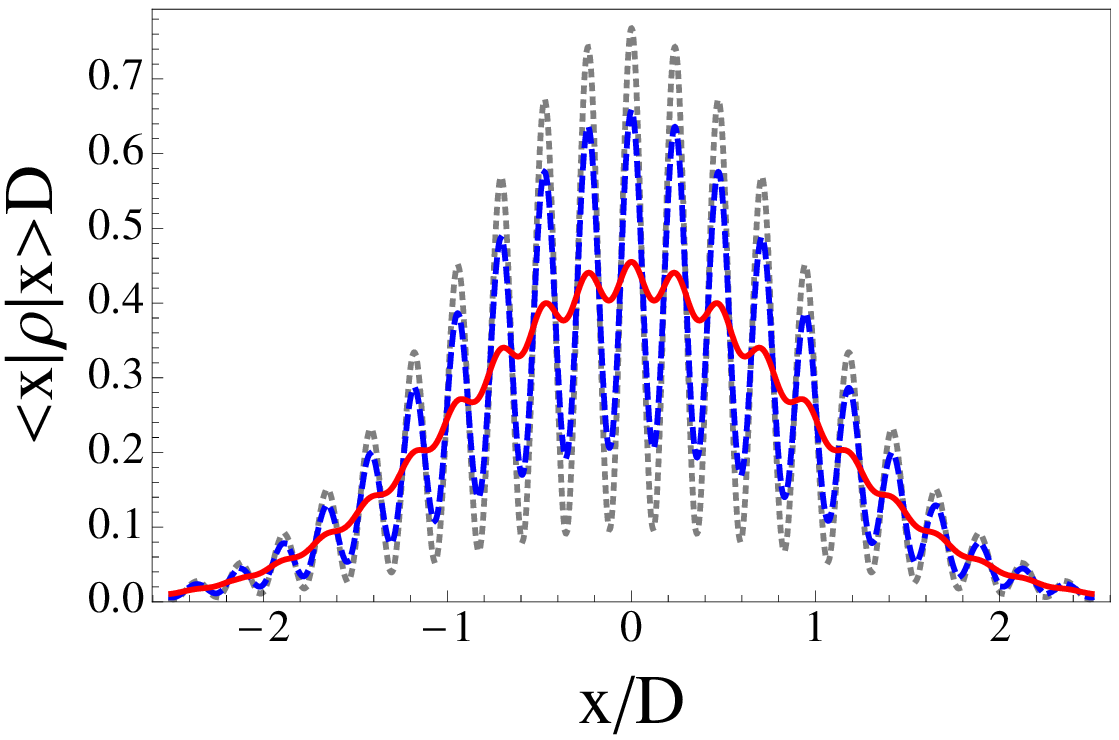}
\caption{(color online). Operational parameter regime of \acron~where the CSL and the QG collapse model can be tested. The upper two panels show the $d$ vs. $D$ diagram as in Fig.~\ref{Fig:Limitationsond} with $P=10^{-14}$ Torr, $T_{i}=100$ K, and $\chi=1000$. The green region is the non-overlapping allowed region taking into account only standard decoherence and the CSL (first panel) or QG (second panel) collapse model. The dot is at $d=30$ nm and $D=100$ nm. For this value, the third lower panel plots the simulated interference pattern taking into account standard decoherence only (dotted gray line), the CSL model (dashed blue line) and the QG model (solid red line).}
\label{Fig:TestCSLQG}
\end{center}
\end{figure}

We start with the stronger collapse models: the CSL and the QG. We will use the $d$ vs.~$D$ diagram (Fig.~\ref{Fig:Limitationsond}) to determine how they can be falsified using \acron.  From the bounds listed in  Table~\ref{t:summary}, the  decoherence given by the collapse model contributes only to conditions (v) $d< \xi(t_{1})$ and (vii) $d < \sqrt{3/(\Lambda t_{2})}$.  Hence, we recalculate these two bounds using $\Lambda=\Lambda_{\text{bb}} + \Lambda_{\text{CM}}$, where $\Lambda_{\text{CM}}$ is the contribution given by the collapse model; this is summarized in Table~\ref{t:parametersCM}.  The times $t_{1}$ and $t_{2}$ depend only on the standard decoherence and  are thus chosen as described in the caption of Fig.~\ref{Fig:Limitationsond}. Figure \ref{Fig:TestCSLQG} shows the $d$ vs.~$D$ diagram including either the CSL or the QG collapse model. As explained in the caption of the figure, the green region is the parameter regime where the collapse model can be falsified. This shows that both the CSL and the QG collapse model can be tested, for instance, at $P=10^{-14}$ Torr, $T_{i}=100$ K, and $\chi=1000$, for spheres with a diameter ranging from 100 nm to 500 nm approximately. As expected, this region is larger for the QG model than for the CSL model with $\gamma^{0}_{\text{CSL}}=10^{-16}$ Hz. The appearance of the green region is mainly due to the tighter bound given by condition (vii) $d < \sqrt{3/(\Lambda t_{2})}$ which is imposed in order to preserve the visibility of the interference pattern. A simulation of the interference pattern is also plotted in Fig.~\ref{Fig:TestCSLQG} for a particular point in the diagram, see the figure's caption. The simulation is done  by numerically solving the master equation at the different steps of \acron. Therefore, both the CSL (with the conservative value of $\gamma^{0}_{\text{CSL}}=10^{-16}$ Hz) and the QG collapse model can be falsified with the successful observation of the interference pattern if \acron~were implemented at the the green region of the parameter regime. 

These results are in strong contrast with the ones obtained for the much weaker exotic decoherence given by the DP and K collapse models. Indeed,  the green region is zero for both cases even at much higher vacuum conditions and low bulk temperatures. However, we have shown in Fig.~\ref{Fig:CollapsevsBB} that the localization parameter of these collapse models is larger than the one given by blackbody radiation at $T_{i}=4.5$ K, and thus their additional exotic decoherence should be observable. However, the problem in this case is that the time required to observe this decoherence using \acron~is longer than the coherence time allowed by the scattering of air molecules $1/\gamma_{\text{air}}$ (see Fig.~\ref{Fig:Figair}). More specifically, the reduction in the visibility for $d \ll 2 a_\text{CM}$ and $t_{2} \ll 1/\gamma_{\text{air}}$ is given by
\be
\mathcal{V}(t_{2})=\exp \left[ - (\Lambda_{\text{bb}}+\Lambda_{\text{CM}}) d^{2} t_{2}/3 \right].
\ee
Assuming $\Lambda_{\text{bb}} \ll \Lambda_{\text{CM}}$, the visibility can be approximated to $\mathcal{V}(t_{2}) \approx \exp \left[ -  \Lambda_\text{CM} d^{2} t_{2}/3 \right]$, and thus it will be reduced at times
\be
  \frac{3}{\Lambda_{\text{CM}} d^{2}} \lesssim t_{2} \ll \frac{1}{\gamma_{\text{air}}},
\ee
where we added the second inequality to emphasize that $t_{2}$ has to be smaller than the coherence time allowed by scattering of air molecules. By inspection of Figs.~\ref{Fig:Figair}, \ref{Fig:PotentialDP}, and \ref{Fig:PotentialK}, one realizes that these two conditions are extremely challenging to be fulfilled for the DP and K model. Therefore, we conclude that the DP and the K model cannot be tested using the present form of \acron. However, the fact that the localization parameter of these collapse models is larger than the one given by blackbody radiation at cryogenic bulk temperatures, hints at the fact that an improved version of \acron, in which, for instance, the free dynamics are accelerated by using repulsive potentials, could meet this challenge.

\begin{table*}[t]
\begin{center}
\begin{ruledtabular}
\begin{tabular}{ |c|c|c|c|c|c|c|c|c|c|c|c| }  
$\rho$ 	&$\epsilon_{r}$				&$\omega$ 					&$\bar n$		&$T_{e}$ 		&$m_{a}$ 		&$\epsilon_{\text{bb}}$ & $\delta x$& $\mathcal{F}$		& $L$  				&$\lambda$  &$W_{c}$  			
\\ \hline 
2201	Kg/$\text{m}^{3}$				&$2.1+\im 10^{-10}$			&$2\pi \times 100 $ KHz					&0.1			&4.5	K			&28.97 amu			&$2.1+\im 0.57 $ & $0.1$ nm &$1.3 \times 10^{5}$	&$2$ $\mu$m		&$1064$ nm & $1.5$ $\mu$m 	
 \end{tabular}
\end{ruledtabular}
\caption{\label{t:parameters} Experimental parameters used in this article. }
\end{center}
\vspace{-0.6cm}
\end{table*}

\section{Optomechanical Double Slit} \label{sec:OMdoubleslit}

The last section of this article is devoted to the analysis of the restrictions that an optomechanical implementation of the squared position measurement imposes on \acron. The implementation of \acron~using cavity optomechanics with levitating dielectric spheres has been recently proposed in~\cite{Romero-Isart2011c}. Here, we provide a thorough derivation and we better remark the conditions that need to be fulfilled. For further literature on cavity optomechanics with levitating dielectrics, we refer the reader to \cite{Romero-Isart2011, Romero-Isart2010b, Chang2010,Romero-Isart2011c}. 

In the following we focus on step \ref{step4} of the protocol, where the sphere is assumed to enter a small cavity, ideally aligned such that the mean position along the cavity axis $\avg{\xop}$ of the sphere is at the node of a cavity mode. In this configuration, the optomechanical coupling is quadratic with $\xop$. This implies that the output light of the cavity contains information about $\xop^{2}$, and therefore, this can be measured by homodyning the light. The optomechanical Hamiltonian reads~\cite{Romero-Isart2011}%
\be \label{eq:HOM}
\hat H(t)= \frac{\pop^{2}}{2m} + \hbar \bar g \adop \aop \xtild^{2} + \im \hbar E(t) (\aop - \adop).
\ee
The first term describes the kinetic energy of the sphere along the cavity axis (note that that there is no harmonic potential since the particle does not need to be trapped during the short interaction required to measure $\xop^{2}$). The third term describes a time dependent driving at frequency $\omega_{L}$ which equals to the cavity resonant frequency $\omega_{c}$, which is used to parametrize the short light pulse. Finally, the second one is the important term describing the optomechanical coupling when the sphere is placed at the node of the cavity mode. 
We have defined the creation (annihilation) operators of the cavity modes $\adop$ ($\aop$), the dimensionless position operator $\tilde x = \xop/\sigma$, and the optomechanical coupling rate given by $\bar g= g_{0} \sigma^{2}/x_{0}^{2}$, where
\be
g_{0}=\epsilon_c  x^{2}_0  k_c^3 c  \frac{V}{4 V_c}
\ee
in the case of a nanosphere~\cite{Romero-Isart2011c, Romero-Isart2011,Romero-Isart2010b, Chang2010}. Here, $\epsilon_c \equiv 3 \text{Re} \left[ (\epsilon_r-1)/(\epsilon_r+2) \right]$ depends on the relative dielectric constant $\epsilon_r$,  $k_{c}= \omega_{c}/c$, and $V_c=\pi W^{2}_{c} L /4$ is the cavity volume, where $W$ is the waist of the cavity mode and $L$ the length of the cavity. As discussed in  \cite{Romero-Isart2011c}, note that $\bar g$ enhances $g_{0}$ by a potentially very large factor $\sigma^{2}/x_{0}^{2}$ depending on the size of the wave packet. The interaction time is assumed to be very small so that the interaction is in the regime of pulsed optomechanics~\cite{Vanner2010a}. We do not take into account the optimization of the pulse shape~\cite{Vanner2010a} and simply consider
a time dependent driving frequency given by $E(t) = \sqrt{2 \kappa n_{\text{ph}}} \xi(t)$, where $\xi(t)$ is a flat-top function of length $T$ and amplitude $ \sim 1/\sqrt{T}$ such that $\int_{0}^{T} \xi^{2}(t) dt=1$, $\kappa$ is the decay rate of the cavity, and $n_{\text{ph}}$ is the total number of photons that the light pulse carries. The decay rate of the cavity has a contribution given by the finesse $\mathcal{F}$ of the empty cavity and by light scattering, and it reads~\cite{Romero-Isart2011c,Chang2010}
\be
\kappa= \frac{2 \pi}{2 \mathcal{F} L} + \frac{c \epsilon_{c}^{2} V^{2} k^{4}_{c}}{16 \pi V_{c}}.
\ee

\subsection{Measurement operator and strength}
Let us show here that after a short interaction with the light pulse, the measurement of the phase quadrature of the output light realizes a measurement of $\xop^{2}$ with some given measurement strength $\chi$. Pulsed optomechanics~\cite{Vanner2010a} consists in implementing a very short interaction time $T\sim \kappa^{-1}$ such that
\be
\frac{\avg{\pop^{2}}}{2m} \frac{T}{\hbar} = \frac{(2 \bar n +1)\omega T}{4 }   \ll 1.
\ee
This allows us to neglect the kinetic term in \eqcite{eq:HOM}, which yields
\be
\hat H(t) \approx \hbar \bar g \adop \aop \xtild^{2} - \im \hbar E(t) (\adop - \aop).
\ee
In order to obtain the output light quadrature we make use of the input-output formalism~\cite{Gardiner2004}. The Langevin equation associated to $\aop$ is given by
\be
\dot a (t)= - (\im \bar g \xtild^{2}+\kappa) \aop(t) + E(t) + \sqrt{2 \kappa} \ain(t),
\ee
where $\ain$ is the input cavity noise operator. We further assume that $\kappa \gg \bar g$, such that one can adiabatically eliminate the cavity mode by setting $\dot a(t)=0$. This leads to
\be
\aop(t) \approx ( E(t) + \sqrt{2 \kappa} \ain(t)) \left( \frac{1}{\kappa} - \frac{\im \bar g \xtild^{2}}{\kappa^{2}}\right).
\ee
By using the input output relation $\aout(t)=\sqrt{2 \kappa} \aop(t) - \ain(t)$ and defining the phase quadrature $\Pout(t) \equiv \im (\adout(t)-\aout(t))/\sqrt{2}$, one obtains the relation
\be
\Pout(t) \approx \Pin(t) + \chi(t) \xtild^{2},
\ee
where $\chi(t) \equiv 2 \bar g E(t)/(\kappa\sqrt{\kappa})$, and we have neglected the small term $\sim  2 \bar g \tilde x^ {2}\Xin /\kappa$. A balanced homodyne measurement of the output field performs a quantum measurement of the time-integrated output quadrature given by~\cite{Romero-Isart2011c}
\be
\Pout \equiv \frac{1}{\sqrt{T}} \int_{0}^{T} \Pout(t) dt = \Pin + \chi \xtild^{2}.
\ee
An important result is the value of the measurement strength, which is given by
\be \label{eq:chiOM}
\chi \approx2 \sqrt{2} \frac{ \bar g \sqrt{n_{\text{ph}}}   }{\kappa}.
\ee
An optimization of the pulse shape provides a different pre-factor which slightly increases the measurement strength (see~\cite{Vanner2010a,Vanner2011}). If the measurement of the optical phase yields the measurement outcome $p_L$, the measurement operator describing the collapse of the center-of-mass state of the sphere is given by~\cite{Romero-Isart2011c, Vanner2011,Vanner2010a}
\be \label{eq:measurement}
\hat{\mathcal{M}} = \exp{ \left[  \im \phi_\text{ds} \tilde x^2 -\left( x_L - \chi \tilde x^2 \right)^2\right]}.
\ee
As a result of the measurement operator of  Eq.~\eqref{eq:measurement}, a superposition of two wave  packets, separated by a distance $d=2 \sigma \sqrt{x_L/\chi}$ and a width given by approximately $\sigma_d \sim \sigma /(4\sqrt{x_L \chi}) = \sigma^2/(2 d \chi)$, is prepared. This is thus in full agreement with the treatment of~Sec.~\ref{sec:DS}. Furthermore, the global phase accumulated during the interaction   with the classical part of the field is given by
\be \label{eq:OMphase}
\begin{split}
\phi_{\text{ds}}&= -\int_{0}^{T} \bar g \avg{\adop(t) \aop(t)} dt\\
& = -\int_{0}^{T} E^{2}(t) \left( \frac{1}{\kappa^{2}} + \frac{\bar g^{2} \avg{\xtild^4}} {\kappa^{4}} \right) dt \approx -\frac{2 \bar g n_{\text{ph}}}{\kappa},
\end{split}
\ee
where the second term can be neglected in the regime $\kappa \gg \bar g$.

\subsection{Restrictions to \acron}

The double slit implementation imposes the following restriction on \acron. First, recall that the phase (\eqcite{eq:OMphase}) needs to be compensated by the phase accumulated during the time of flight $\phi_{\text{tof}}=t_{1} \omega /4$. Thus, $\phi_{\text{tof}}+ \phi_{\text{ds}} \sim 0$ is fulfilled when the total number of photons in the light pulse is 
\be \label{eq:eqphase}
n_{\text{ph}} =\frac{\omega t_{1} \kappa }{8 g_{0}} \frac{x^{2}_{0}}{\sigma^2} \approx \frac{ \kappa }{8 g_{0} t_{1} \omega}.
\ee
Here we have used again that $\sigma^{2} \approx x^{2}_{0} t^{2}_{1} \omega^{2}$ at  times $t_{1} \omega \gg 1$. Inserting equality \eqcite{eq:eqphase} into the definition of the measurement strength \eqcite{eq:chiOM} leads to
\be \label{eq:chiOMt}
\chi \approx \left(t_{1} \omega \right)^{3/2} \sqrt{\frac{g_{0}}{\kappa}}.
\ee

Additionally, standard decoherence due to light scattering during the light-mechanics interaction~\cite{Romero-Isart2011c,Chang2010} is prevented if the following conditions are met. This decoherence is also of the localization type, with a decoherence rate given by $\Gamma_{\text{sc}}(t) = \Lambda^{0}_{\text{sc}} E^{2}(t) \sigma^2/\kappa$ for distances smaller than the optical wavelength, where the localization parameter is~\cite{Romero-Isart2011c,Chang2010}
\be
\Lambda^{0}_{\text{sc}}=\frac{\epsilon^{2}_{c}}{6 \pi} \frac{ c }{ V_{c} }V^{2} k_{c}^{6}.
\ee
This form of decoherence is prevented as long as $\int_{0}^{T}\Gamma_{\text{sc}}(t) dt \ll 1$, which gives rise to the following condition on $t_{1}$
\be
t_{1}   \ll \frac{ 4 g_{0}}{ \omega \Gamma^{0}_{\text{sc}}},
\ee
where we have used \eqcite{eq:eqphase} and we have defined $\Gamma^{0}_{\text{sc}} \equiv\Lambda^{0}_{\text{sc}}  x_{0}^{2} $. Bear in mind that the adiabatic elimination used in the derivation is valid as long as $\kappa \gg \bar g = g_{0} t^{2}_{1} \omega^{2}$, which leads a further constrain, namely $t_{1} \ll \omega^{-1} \sqrt{\kappa/g_{0}}$. Thus, the optomechanical implementation of the step \ref{step4} of \acron~yields an additional upper bound on $t_{1}$ given by 
\be  \label{eq:OMt1}
t_{1} \ll \frac{1}{\omega} \min \left\{  \sqrt{\frac{\kappa}{g_{0}}}, \frac{ 4 g_{0}}{  \Gamma^{0}_{\text{sc}}}  \right\}\equiv t_{1}^{\text{OM}}.
\ee
This is incorporated in  Table~\ref{t:summary} as condition (ix).  Also, by inserting \eqcite{eq:OMt1} into \eqcite{eq:chiOMt} we obtain an upper bound for the measurement strength given by
\be \label{eq:chiOM}
\chi \ll  \min \left\{  \left(\frac{\kappa}{g_{0}} \right)^{1/4}, \frac{8g^{2}_{0}}{\sqrt{\kappa   (\Gamma^{0}_{\text{sc}})^3 }   }\right\}\equiv \chi_{\text{max}}.
\ee

\begin{figure}
\begin{center}
\includegraphics[width=\linewidth]{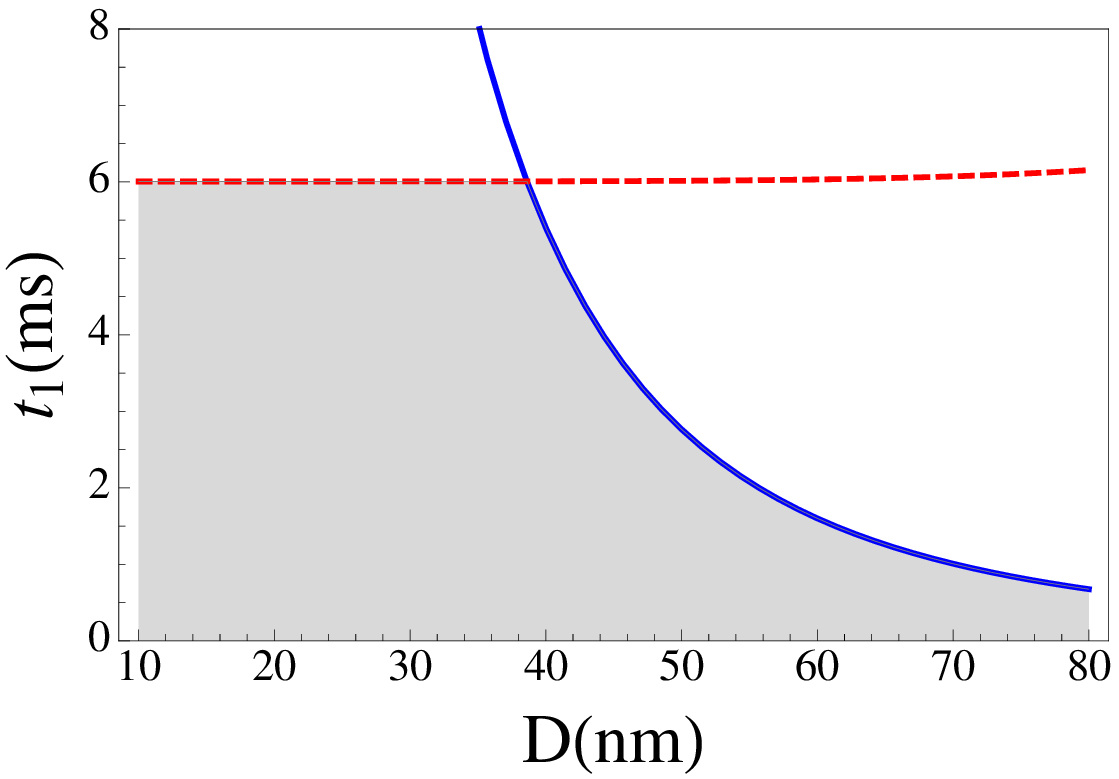}
\includegraphics[width=\linewidth]{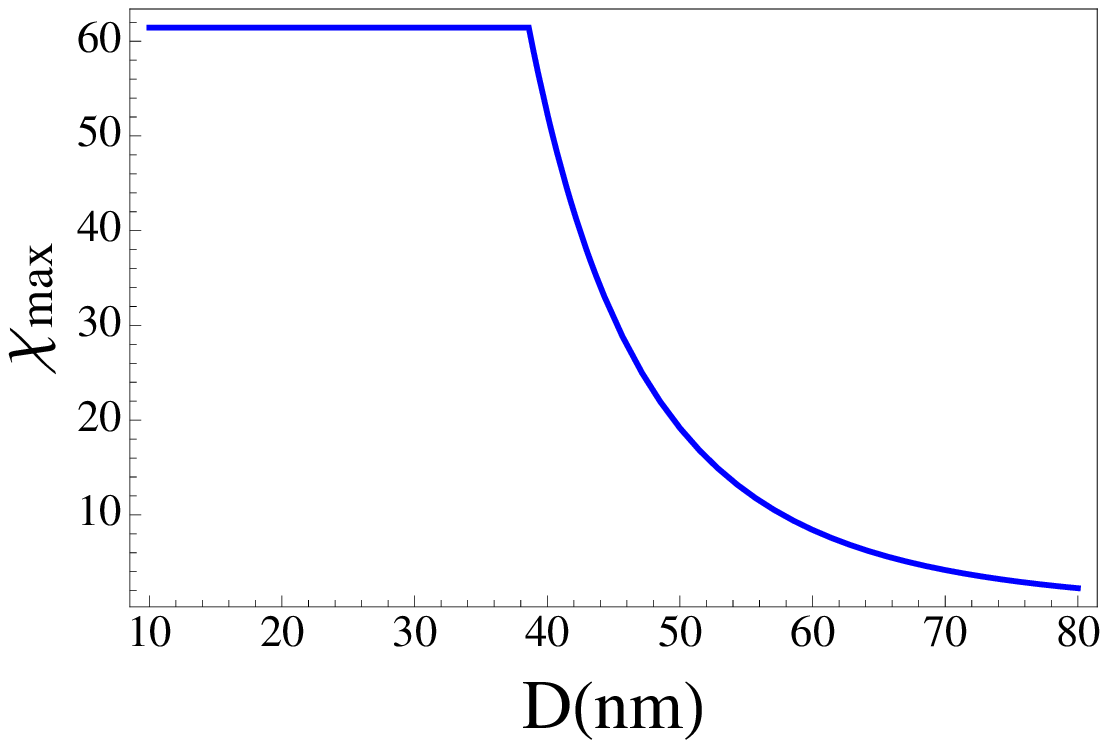}
\caption{(color online). In the upper panel, $t_{1}$ (see \eqcite{eq:OMt1}) is plotted as a function of the diameter of the sphere. The solid blue line is the upper bound $4 g_{0} /(\omega \Gamma_{\text{sc}}^{0})$ and the dashed red line the $\omega^{-1}(\kappa/g_{0})^{1/4}$, which is the bound given by the adiabatic condition. The gray region shows allowed $t_{1}$ values. In the lower panel, $\chi_{\text{max}}$ (see \eqcite{eq:chiOM}) is plotted. We used a fiber-based Fabry-Perot optical cavity, see main text and Table~\ref{t:parameters}. }
\label{Fig:t1chiOM}
\end{center}
\end{figure}

\begin{figure}[t]
\begin{center}
\includegraphics[width=\linewidth]{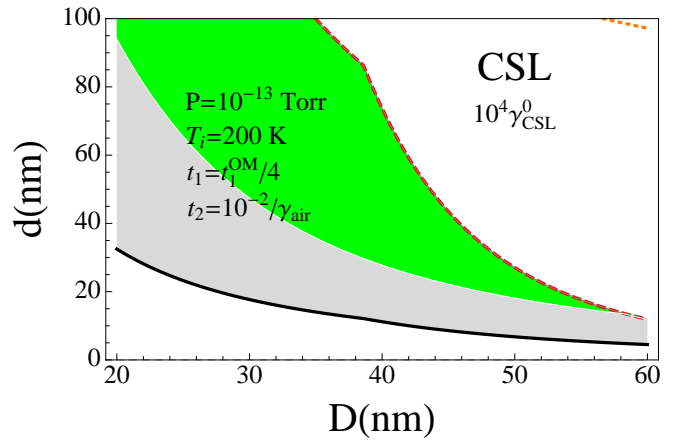}
\caption{(color online). The $d$ vs.~$D$ diagram (see Fig.~\ref{Fig:Limitationsond}) is plotted taking into account the additional limitations imposed by an optomechanical implementation of the squared position measurement. We used the experimental parameters of Table~\ref{t:parameters}, a pressure of $P=10^{-13}$ Torr, an internal bulk temperature of $T_{i}=200$ K, $t_{1}=t_{1}^\text{OM}/4$, and $t_{2}=10^{-2}/\gamma_{\text{air}}$. }
\label{Fig:TestCSLOM}
\end{center}
\end{figure}

These conditions give a strong limitation for the overall performance of \acron~which crucially depend on the quality of the optical cavity employed. In~\cite{Romero-Isart2011c}, it was suggested to use the recently developed fiber-based Fabry-Perot cavities of length of 2 $\mu$m and finesse $\mathcal{F} \approx 1.3 \times 10^{5}$~\cite{Hunger2010}. As shown in Fig.~\ref{Fig:t1chiOM}, for this cavity, upper bounds for $t_{1}$ of the order of milliseconds and a corresponding $\chi_{\text{max}}$ of several tens are obtained for spheres smaller than 100 nm. To see the implications for the realization of \acron, we plot in Fig.~\ref{Fig:TestCSLOM} the $d$ vs.~$D$ diagram for a pressure of $P=10^{-13}$ Torr and a bulk temperature of $T_{i}=200$ K (which is reasonable considering the heating produced by laser absorption~\cite{Chang2010,Romero-Isart2011c}). For these parameters, $t_{1}=t_{1}^{\text{OM}}/4$ guarantees the fulfillment of the  bounds on $t_{1}$ given in Table~\ref{t:parameters}. It is remarkable that even taking into account the restrictions imposed by the optomechanical implementation, spheres with a diameter of tens of nanometers that contain of the order of $10^{7}$ atoms can be prepared in superpositions of the order of their size. Moreover, even the CSL model with a localization rate frequency given by $10^{4} \gamma^{0}_{\text{CSL}}$, which is orders of magnitudes lower than the enhancement predicted by Adler~\cite{Adler2007,Adler2009}, can be falsified.  The result shown in Fig.~\ref{Fig:TestCSLOM} is very similar to the one given in~\cite{Romero-Isart2011c}. Note however that here we have used a pressure three orders of magnitude larger due to the saturation effect omitted in~\cite{Romero-Isart2011}. This renders the implementation of \acron~using cavity optomechanics with levitating spheres less challenging.  Finally, we remark the recent proposal given in~\cite{Nimmrichter2011b} to test the CSL model using an all-optical time-domain Talbot-Lau interferometer for clusters with masses exceeding $10^6$ amu.

\section{Conclusions}\label{sec:conclusions}

In summary, we have shown that by combining techniques and insights from quantum-mechanical resonators and matter-wave interferometry, one can prepare large spatial quantum superpositions of massive objects comparable to their size. The protocol consists of cooling a mechanical resonator to its ground state, switching off the harmonic potential to let the wave function coherently expand, preparing a spatial quantum superposition by performing a measurement of the squared position observable, and observing interference by measuring the position after further free evolution. We have focused on solid spheres with diameters ranging from tens of nanometers to few micrometers. We have taken into account unavoidable sources of decoherence such as the scattering of environmental massive particles and the emission, absorption, and scattering of blackbody radiation. Both sources provide coherence times of the order of milliseconds within reasonable values for pressures and temperatures. At low pressures, decoherence due to blackbody radiation is dominant when the bulk temperature is larger than the cryogenic environmental temperature. Additional limitations are given by the slow free dynamics involved for these massive objects. For larger masses, the wave function  takes longer to coherently expand and to build a visible interference pattern after a superposition has been prepared. 

We have also argued that this protocol can be  applied to test some of the most paradigmatic collapse models. In particular, we have analyzed the continuous spontaneous localization model (CSL), a model based on quantum gravity (QG), the Di\'osi-Penrose model (DP), and the K\'arolyh\'azy model (K). The CSL and the DP are much stronger than the DP and  K model and can be falsified using reasonable experimental parameters. In particular, the famous continuous spontaneous localization model can be tested using the original and conservative choice of parameters. However, the DP and K model are much more challenging to falsify despite the fact that they predict a decoherence which is stronger than the one provided by standard decoherence at low bulk temperatures. Nevertheless, these models are strongly limited by the fact that the free dynamics of the large masses required is too slow. We remark that for the Di\'osi-Penrose model we did not consider the strong enhancement provided by taking into account the mass density at the microscopic level. The later, besides being a controversial choice, turns the model into a one parameter model given by the mass resolution parameter $r_{0}$. Note that if this parameter is taken into account, the protocol proposed here provides unprecedented lower bounds to its value. 

We have also addressed the optomechanical implementation of the protocol presented, namely \acron. We focused on the squared position measurement required to perform the double slit, and we have considered cavity optomechanics with optically levitating nanospheres.  We have shown that the overall performance of the protocol is limited by this implementation, since both the global phase added during the interaction and light scattering set upper bounds on the expansion time and the measurement strength.  Nevertheless, with recently developed fiber-based cavities and for spheres of the order of tens of nanometers, superpositions of the order of their size could be prepared. This  provides unprecedented bounds to the continuous spontaneous localization model that can be used to falsify the enhancement of the localization rate predicted by Adler.

There are various directions to further pursue the work presented here. First,  the study of the implementation of \acron~using cavity optomechanics with suspended disks~\cite{Chang2010a}. In this setup, the mechanical frequency can  also be varied since the tight harmonic potential is achieved by optical trapping, but the scattering of light is strongly reduced when the laser waist is smaller than the disk. This comes at the price of inducing decoherence due to the coupling with internal elastic modes. Second, the possibility of using repulsive potentials to exponentially increase the time scales of the free dynamics, and thus, to efficiently use the long coherence times given by scattering of air molecules and blackbody radiation. Naturally, this has to be done without incorporating additional sources of decoherence. In any case, we believe that the synergy between the fields of quantum-mechanical resonators and matter-wave interferometry will allow to explore in the near future the limits of quantum mechanics at unprecedented scales, an exciting possibility indeed.

I am grateful to J. I. Cirac, A. C. Pflanzer, F. Blaser, R. Kaltenbaek, N. Kiesel, and M. Aspelmeyer for useful discussions. I acknowledge support from the Alexander von Humboldt Stiftung.

\end{document}